\documentclass{elsart}
\usepackage{ifpdf}
\usepackage{graphicx,amssymb,lineno,epsfig,subfig}
\usepackage{subfig}
\usepackage{indentfirst}
\ifpdf
\usepackage[%
  pdftitle={Instructions for use of the document class
    elsart},%
  pdfauthor={Simon Pepping},%
  pdfsubject={The preprint document class elsart},%
  pdfkeywords={instructions for use, elsart, document class},%
  pdfstartview=FitH,%
  bookmarks=true,%
  bookmarksopen=true,%
  breaklinks=true,%
  colorlinks=true,%
  linkcolor=blue,anchorcolor=blue,%
  citecolor=blue,filecolor=blue,%
  menucolor=blue,pagecolor=blue,%
  urlcolor=blue]{hyperref}
\else
\usepackage[%
  breaklinks=true,%
  colorlinks=true,%
  linkcolor=blue,anchorcolor=blue,%
  citecolor=blue,filecolor=blue,%
  menucolor=blue,pagecolor=blue,%
  urlcolor=blue]{hyperref}
\fi

\renewcommand{\theequation}{\arabic{section}.\arabic{equation}}
\newtheorem{theorem}{Theorem}[section]
\newtheorem{remark}{Remark}[section]
\newtheorem{example}{Example}[section]

\def\@subjclass{}
\makeatletter
\def\elsartstyle{%
    \def\normalsize{\@setfontsize\normalsize\@xiipt{14.5}}
    \def\small{\@setfontsize\small\@xipt{13.6}}
    \let\footnotesize=\small
    \def\large{\@setfontsize\large\@xivpt{18}}
    \def\Large{\@setfontsize\Large\@xviipt{22}}
    \skip\@mpfootins = 18\p@ \@plus 2\p@
    \normalsize
} \@ifundefined{square}{}{\let\Box\square} \makeatother
\usepackage{setspace}

\usepackage{indentfirst}
\usepackage{setspace}

\begin{document}
\begin{frontmatter}
\title{New exact travelling wave solutions for the $K(2,2)$
   equation with osmosis dispersion}
\author{Jiangbo Zhou\corauthref{cor}},
\corauth[cor]{Corresponding author. Tel.: +86-511-88969336; Fax:
+86-511-88969336.} \ead{zhoujiangbo@yahoo.cn}
\author{Lixin Tian},
\author{Xinghua  Fan}
\address{Nonlinear Scientific Research Center, Faculty of Science, Jiangsu
University, Zhenjiang, Jiangsu 212013, China}
\begin{abstract} In this paper, by using bifurcation method, we successfully find the
$K(2,2)$ equation with osmosis dispersion
$u_t+(u^2)_x-(u^2)_{xxx}=0$ possess two new types of travelling wave
solutions called kink-like wave solutions and antikink-like wave
solutions. They are defined on some semifinal bounded domains and
possess properties of kink waves and anti-kink waves. Their implicit
expressions are obtained. For some concrete data, the graphs of the
implicit functions are displayed, and the numerical simulation of
travelling wave system is made by Maple. The results show that our
theoretical analysis agrees with the numerical simulation.
\end{abstract}


\end{frontmatter}
\section{Introduction}
 \setcounter {equation}{0}
In recent years, many nonlinear partial differential equations
(NLPDEs) have been derived from physics, mechanics, engineering,
biology, chemistry and other fields. Since exact solutions can help
people know deeply the described process and possible applications,
seeking exact solutions for NLPDEs is of great importance.

In 1993, Rosenau and Hyman \cite {1} introduced a genuinely
nonlinear dispersive equation, a special type of KdV equation, of
the form

\begin{equation}
\label {eq1.1}u_t +a(u^n)_x + (u^n)_{xxx}=0, n>1,
\end{equation}
where $a$ is a constant and both the convection term $(u^n)_x$ and
the dispersion effect term $(u^n)_{xxx}$ are nonlinear. These
equations arise in the process of understanding the role of
nonlinear dispersion in the formation of structures like liquid
drops. Rosenau and Hyman derived solutions called compactons for
Eq.(\ref{eq1.1}) and showed that while compactons are the essence of
the focusing branch where $a>0$, spikes, peaks, and cusps are the
hallmark of the defocusing branch where $a<0$  which also supports
the motion of kinks. Further, the negative branch, where $a<0$, was
found to give rise to solitary patterns having cusps or infinite
slopes. The focusing branch and the defocusing branch represent two
different models, each leading to a different physical structure.
Many powerful methods were applied to construct the exact solutions
for Eq.(\ref{eq1.1}), such as Adomain method \cite {2}, homotopy
perturbation method \cite {3}, Exp-function method \cite {4},
variational iteration method \cite {5}, variational method \cite {6,
7}. In \cite {8}, Wazwaz studied a generalized forms of the
Eq.(\ref{eq1.1}), that is $mK(n,n)$ equations and defined by
\begin{equation}
 \label {eq1.2} u^{n-1}u_t +a(u^n)_x+ b(u^n)_{xxx}=0, n>1,
\end{equation}
where $a,b$ are constants. He showed how to construct compact and
noncompact solutions for Eq.(\ref{eq1.2}) and discussed it in higher
dimensional spaces in  \cite {9}. Chen et al.  \cite {10} showed how
to construct the general solutions and some special exact solutions
for Eq.(\ref{eq1.2}) in higher dimensional spatial domains. He et
al. \cite {11} considered the bifurcation behavior of travelling
wave solutions for Eq.(\ref{eq1.2}). Under different parametric
conditions, smooth and non-smooth periodic wave solutions, solitary
wave solutions and kink and anti-kink wave solutions were obtained.
Yan \cite {12} further extended Eq.(\ref{eq1.2}) to be a more
general form
\begin{equation}
 \label {eq1.3} u^{m-1}u_t +a(u^n)_x+ b(u^k)_{xxx}=0, nk\neq 1,
\end{equation}
And using some direct ansatze, some abundant new compacton
solutions, solitary wave solutions and periodic wave solutions of
Eq.(\ref{eq1.3}) were obtained. By using some transformations, Yan
\cite {13} obtained some Jacobi elliptic function solutions for
Eq.(\ref{eq1.3}). Biswas \cite {14} obtained 1-soliton solution of
equation with the generalized evolution term
 \begin{equation}
 \label {eq1.4} (u^l)_t +a(u^m)u_x+ b(u^n)_{xxx}=0,
\end{equation}
where $a, b$ are constants, while $l, m$ and $n$ are positive
integers. Zhu et al. \cite {15} applied the decomposition method and
symbolic computation system to develop some new exact solitary wave
solutions for the $K(2, 2, 1)$ equation
\begin{equation}
 \label {eq1.5}  u_t +(u^2)_x - (u^2)_{xxx}+u_{xxxxx}=0,
\end{equation}
 and the $K(3, 3, 1)$ equation
\begin{equation}
 \label {eq1.6}  u_t +(u^3)_x - (u^3)_{xxx}+u_{xxxxx}=0.
\end{equation}
In \cite {16}, Xu and Tian introduced the osmosis $K(2,2)$ equation
\begin{equation}
 \label {eq1.7}  u_t +(u^2)_x - (u^2)_{xxx}=0,
\end{equation}
where the negative coefficient of dispersive term denotes the
contracting dispersion. They obtained the peaked solitary wave
solution and the periodic cusp wave solution for Eq.(\ref{eq1.7}).
In this paper, we'll continue their work and using the bifurcation
method of planar dynamical systems to derive two new types of
bounded travelling wave solutions for Eq.(\ref{eq1.7}). They are
defined on some semifinal bounded domains and possess properties of
kink waves and anti-kink waves. To our knowledge, such type of
travelling wave solution has never been found for Eq.(\ref{eq1.7})
in the former literature.

The remainder of the paper is organized as follows. In Section 2, we
give the phase portrait of the travelling wave system and use Maple
to show the graphs of the orbits connecting with the saddle points
for our purpose. In Section 3, we state the main results which are
implicit expressions of the kink-like  and  antikink-like wave
solutions. In Section 4, we give the proof of the main results. For
some concrete data, we use Maple to display the graphs of the
implicit functions. In Section 5, the numerical simulations of
travelling wave system are made by Maple. A short conclusion is
given in Section 6.

\section{Phase portrait of the travelling wave system}
 \setcounter {equation}{0}

 Eq.(\ref{eq1.7}) also takes the form
\begin{equation}
 \label {eq2.1}  u_t +2uu_x - 6u_xu_{xx}-2uu_{xxx}=0,
\end{equation}

Let $u = \varphi (\xi )$   with $\xi = x - ct (c\neq0)$  be the
solution for Eq.(\ref{eq2.1}), then it follows that
 \begin{equation}
\label{eq2.2}
 - c\varphi' + 2\varphi \varphi '-6\varphi' \varphi '' -2\varphi \varphi
 '''=0.
\end{equation}

Integrating (\ref{eq2.2})  once we have
\begin{equation}
\label{eq2.3} - c\varphi + (\varphi)^2-2( \varphi ')^2-2\varphi
\varphi '' =g,
\end{equation}
\noindent where $g$ is the integral constant.

Let $y = \varphi '$, then we get the following planar dynamical
system:
\begin{equation}
\label{eq2.4} \left\{ {\begin{array}{l}
 \frac{\textstyle d\varphi }{\textstyle d\xi } = y \\
 \frac{\textstyle dy}{\textstyle d\xi } = \frac{\textstyle \varphi ^2-c \varphi-g -2y^2}{\textstyle 2\varphi}\\
 \end{array}} \right.
\end{equation}
\noindent with a first integral
\begin{equation}
\label{eq2.5} H(\varphi, y)=\varphi^2(y^2 -
\frac{1}{4}\varphi^2+\frac{c}{3}\varphi+ \frac{1}{2} g )=h,
\end{equation}
\noindent where $h$ is a constant.

Note that (\ref{eq2.4}) has a singular line $\varphi = 0$, to avoid
the line temporarily we make transformation $d\xi = 2\varphi d\zeta
$. Under this transformation, Eq.(\ref{eq2.4}) becomes
\begin{equation}
\label{eq2.6} \left\{ {\begin{array}{l}
 \frac{\textstyle d\varphi }{\textstyle d\zeta } =  2\varphi y \\
 \frac{\textstyle dy}{\textstyle d\zeta } = \varphi ^2-c \varphi-g -2y^2
\\
 \end{array}} \right.
\end{equation}
Eq.(\ref{eq2.4}) and Eq.(\ref{eq2.6}) have the same first integral
as (\ref{eq2.5}). Consequently, system (\ref{eq2.4}) has the same
topological phase portraits as system (\ref{eq2.6}) except for the
straight line $\varphi = 0$. Obviously, $\varphi = 0$ is an
invariant straight-line solution for system (\ref{eq2.6}).

Now we consider the singular points  of system (\ref{eq2.6}) and
their properties. Note that for a fixed $h$, (\ref{eq2.5})
determines a set of invariant curves of (\ref{eq2.6}). As $h$ is
varied, (\ref{eq2.5}) determines different families of orbits of
(\ref{eq2.6}) having different dynamical behaviors. Let $M(\varphi
_e ,y_e )$ be the coefficient matrix of the linearized system of
(\ref{eq2.6}) at the equilibrium point $(\varphi _e ,y_e )$, then

\[
M(\varphi _e ,y_e ) = \left( {{\begin{array}{*{20}c}
{\indent  y_e } &&& {\indent 2\varphi _e}   \\
 {2\varphi _e - c }   &&& \indent{- 4y_e}   \\
\end{array} }} \right)
\]
\noindent and at this equilibrium point, we have
\[
 J(\varphi _e ,y_e ) = \det M(\varphi _e ,y_e ) = - 4y_e^2 -
4\varphi _e (\varphi _e -\frac {c}{2}\varphi _e ),
\]
\[
  p(\varphi _e ,y_e ) =
\mathrm{trace}(M(\varphi _e ,y_e )) = - 3y_e. \nonumber
\]
By the theory of planar dynamical system (see \cite {17}), for an
equilibrium point of a planar dynamical system, if $J < 0$, then
this equilibrium point is a saddle point; it is a center point if $J
> 0$ and $p = 0$; if $J = 0$ and the Poinc\'{a}re index of the
equilibrium point is 0, then it is a cusp.

Although the distribution and properties of equilibrium points of
system (\ref{eq2.4}) has been given in \cite {16}. Here we also
state it briefly for our purpose. System (\ref{eq2.4}) has the
following properties:

 (1) When $g>0$, system (\ref{eq2.4}) has two
equilibrium points $(\varphi _0^ - ,0)$ and $(\varphi _0^ + ,0)$.
They are saddle points. (i) If $c<0$, then there is inequality
$\varphi _0^ - <\frac{c}{2}<0<\varphi _0^ +$; (ii)  If $c>0$, then
there is inequality $\varphi _0^ -<0 <\frac{c}{2}<\varphi _0^ +$.

 (2) When $g=0$, system (\ref{eq2.4}) has two equilibrium
points $(0,0)$ and $(c,0)$. $(0,0)$ is a cusp, and $(c,0)$ is a
saddle point.

 (3) When $-\frac{c^2}{4}<g<0$, system
(\ref{eq2.4}) has two equilibrium points $(\varphi _0^ - ,0)$ and
$(\varphi _0^ + ,0)$. (i) If $c<0$, then $(\varphi _0^ - ,0)$ is a
saddle point while $(\varphi _0^ + ,0)$ is a center point. There is
inequality $\varphi _0^ - <\frac{c}{2}<\varphi _0^ +<0$; (ii) If
$c>0$ , then $(\varphi _0^ - ,0)$ is a center point while $(\varphi
_0^ + ,0)$ is a saddle point. There is inequality $0<\varphi _0^ -
<\frac{c}{2}<\varphi _0^ +$.

 (4) When $g=-\frac{c^2}{4}$, system
(\ref{eq2.4}) has only one equilibrium point $(\frac{c}{2},0)$. It
is a cusp.

 (5) When
$g<-\frac{c^2}{4}$, system (\ref{eq2.4}) has no equilibrium point.

\begin{remark}
Suppose that $\varphi(\xi )(\xi = x - ct)$ is a travelling wave
solution for Eq.(\ref{eq1.7}) for $\xi\in ( - \infty, + \infty )$,
and $\mathop {\lim }\limits_{\xi \to - \infty } \varphi(\xi ) = A$,
$\mathop {\lim }\limits_{\xi\to\infty } \varphi(\xi ) = B$, where
$A$ and $B$ are two constants. If $A=B$, then $\varphi(\xi )$ is
called a solitary wave solution. If $A \ne B$, then $\varphi(\xi )$
is called a kink (or an anti-kink) solution. Usually, a solitary
wave solution for Eq.(\ref{eq1.7}) corresponds to a homoclinic orbit
of system (\ref{eq2.6}) and a periodic orbit of system (\ref{eq2.6})
corresponds to a periodic travelling wave solution of
Eq.(\ref{eq1.7}). Similarly, a kink (or an anti-kink) wave solution
of Eq.(\ref{eq1.7}) corresponds to a heteroclinic orbit (or
so-called connecting orbit) of
 system (\ref{eq2.6}). In \cite{16}, Xu and Tian reported that when $-\frac{c^2}{4}<g<-\frac{2c^2}{9}$,
system (\ref{eq2.6}) has homoclinic orbits, and when
$-\frac{c^2}{4}<g<0$, system (\ref{eq2.6}) has a periodic orbit
which consist of an arc and a line segment. They obtained peakon
solutions from the limit of solitary waves and from the limit of
periodic cusp waves. We'll obtain a new type of bounded travelling
wave solutions called kink-like and antikink-like wave solutions for
Eq.(\ref{eq1.7}) when $g>-\frac{2c^2}{9}$, which correspond to the
orbits of system (\ref{eq2.6}) connecting with the saddle points.
\end{remark}

We show the phase portraits in each region and on the bifurcation
curves in Fig.1. From Fig.1, we can see that when
$g>-\frac{2c^2}{9}$, system (\ref{eq2.6}) has orbits connecting with
the saddle points.

\begin{figure}[h]
\centering \subfloat[]{\label{fig:1}
\includegraphics[height=1in,width=1.2in]{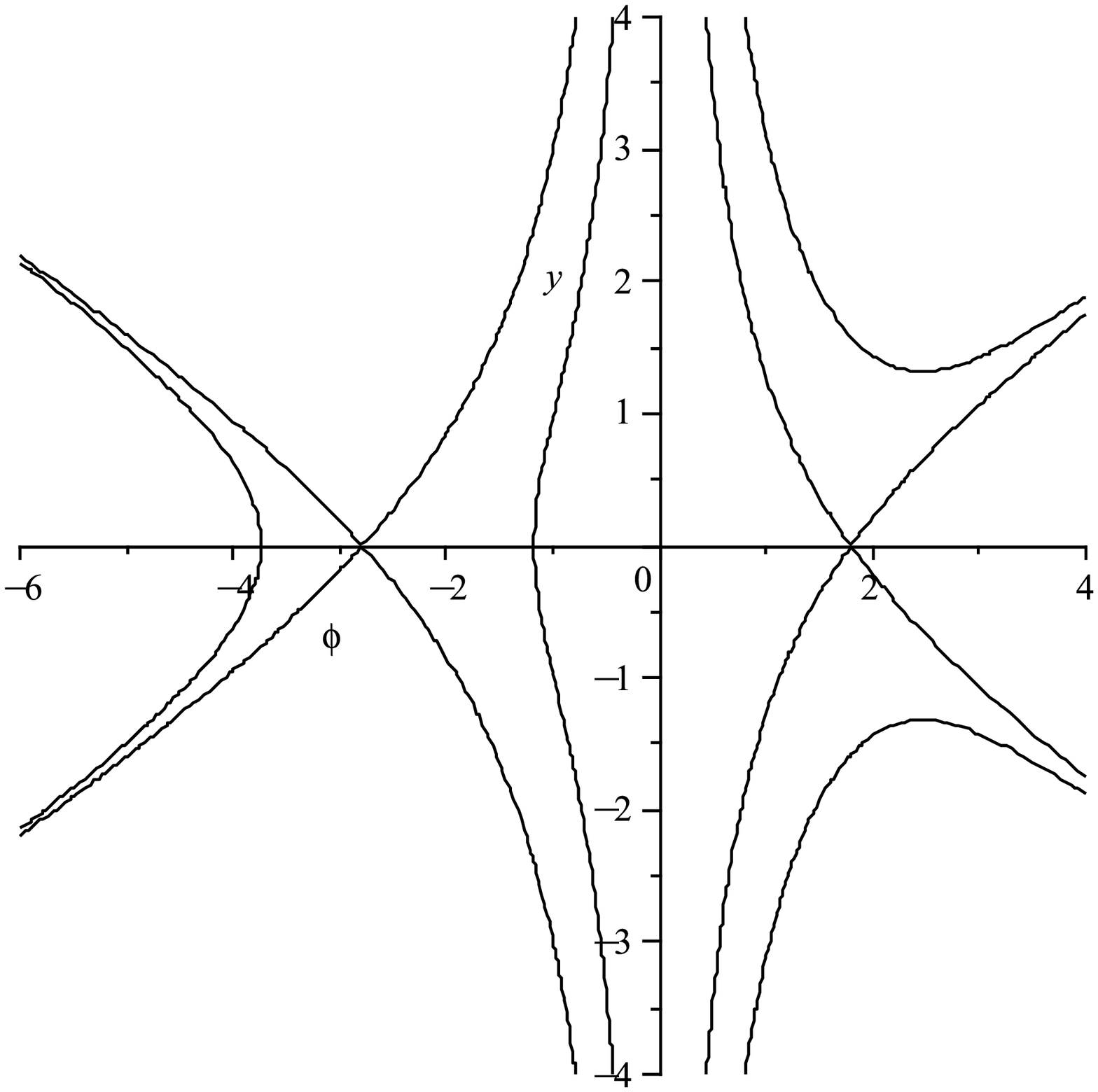}}\hspace{0.01\textwidth}
\subfloat[ ]{ \label{fig:2}
\includegraphics[height=1in,width=1.2in]{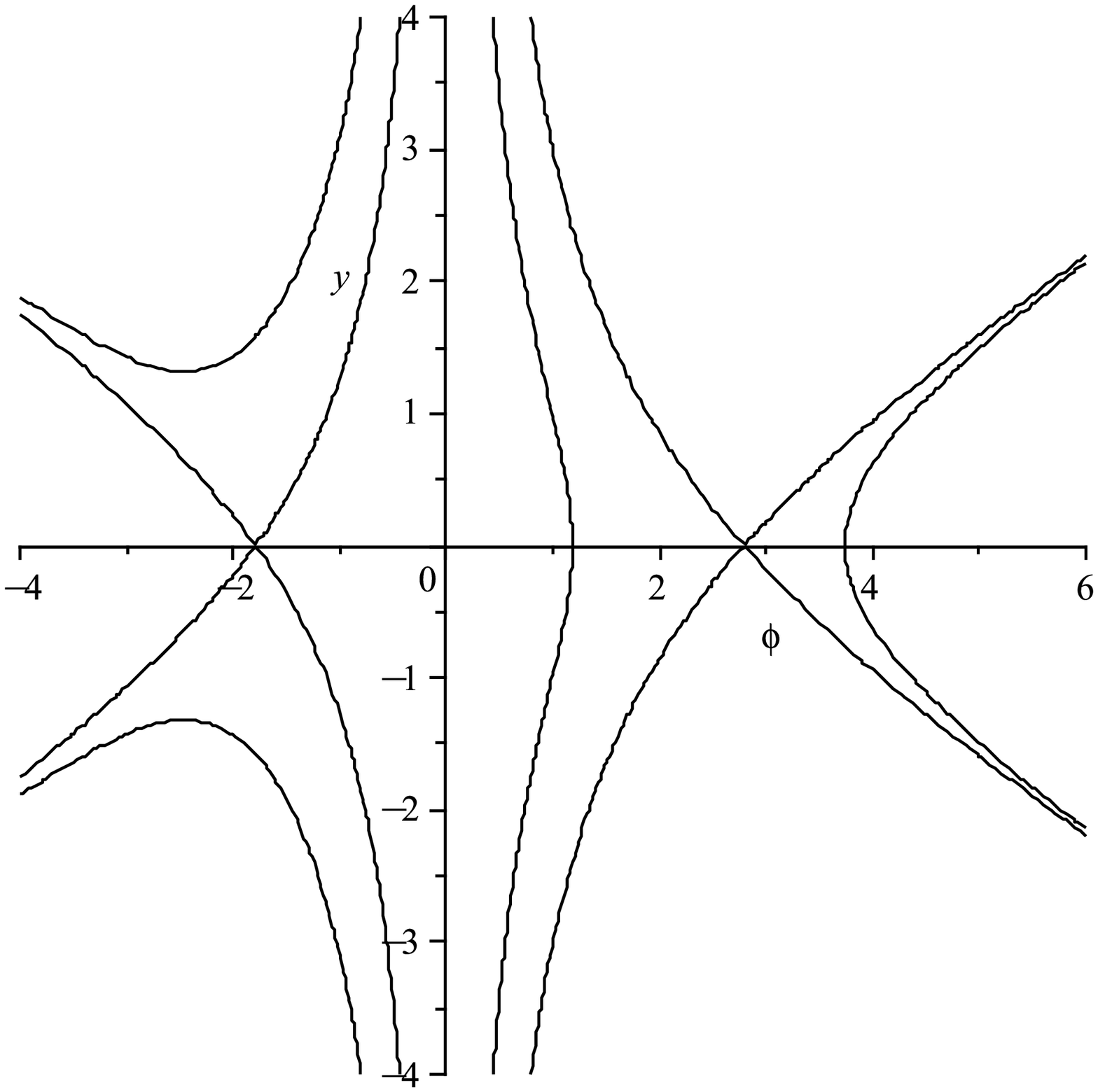}}\hspace{0.01\textwidth}
\subfloat[]{ \label{fig:3}
\includegraphics[height=1.in,width=1.2in]{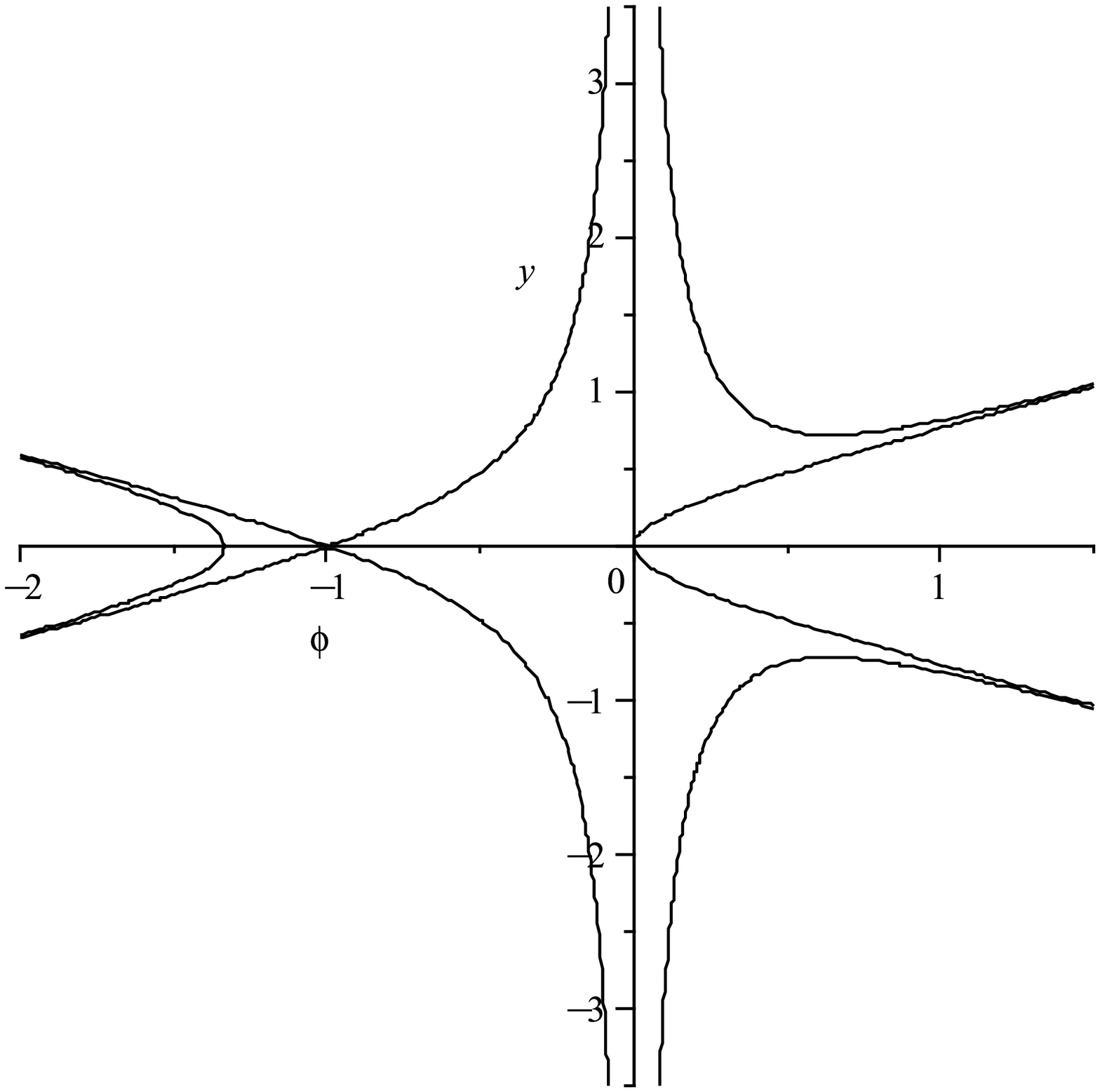}}\hspace{0.01\textwidth}
\subfloat[ ]{ \label{fig:4}
\includegraphics[height=1.in,width=1.2in]{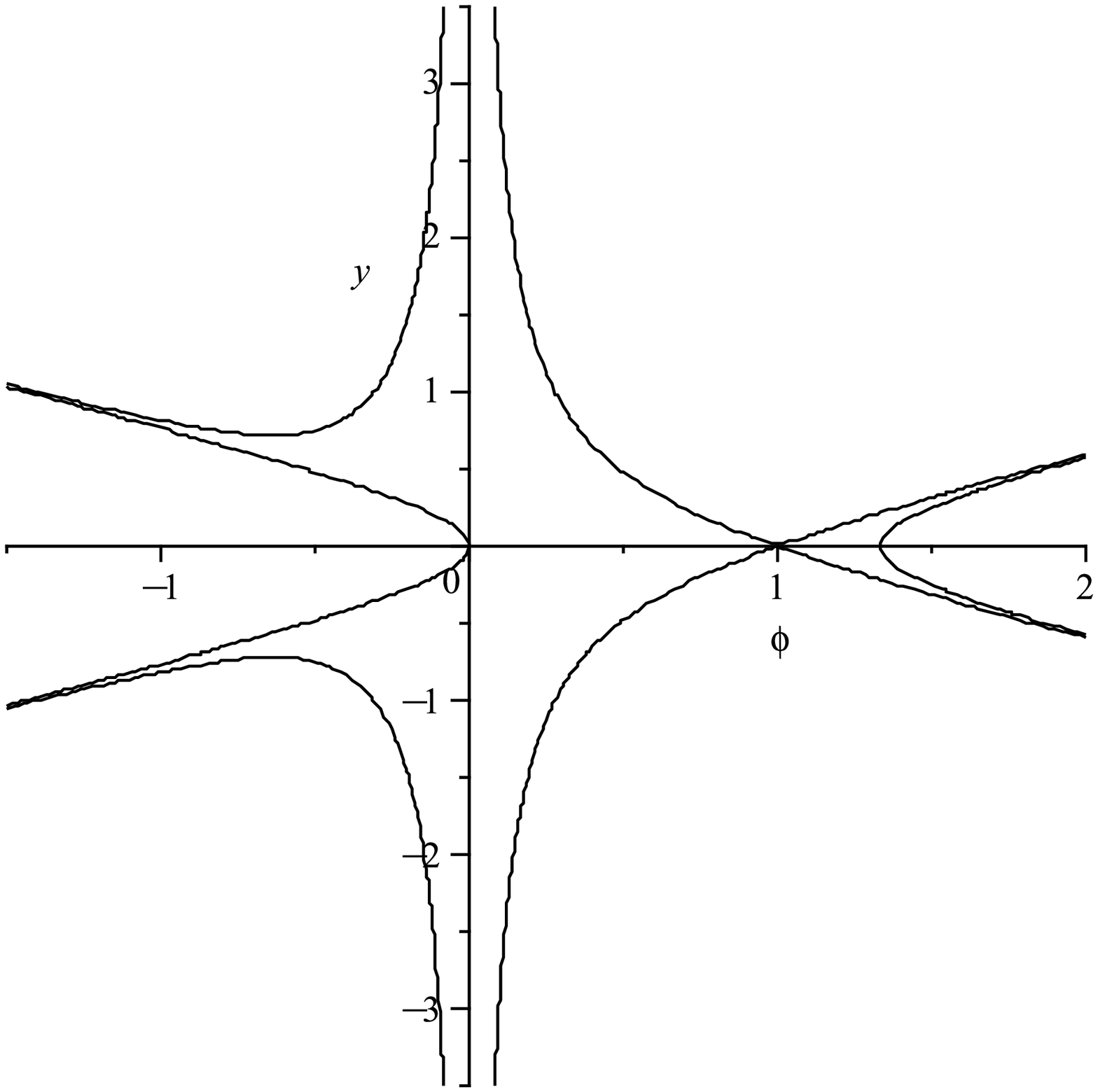}}\\
\subfloat[]{ \label{fig:5}
\includegraphics[height=1.in,width=1.2in]{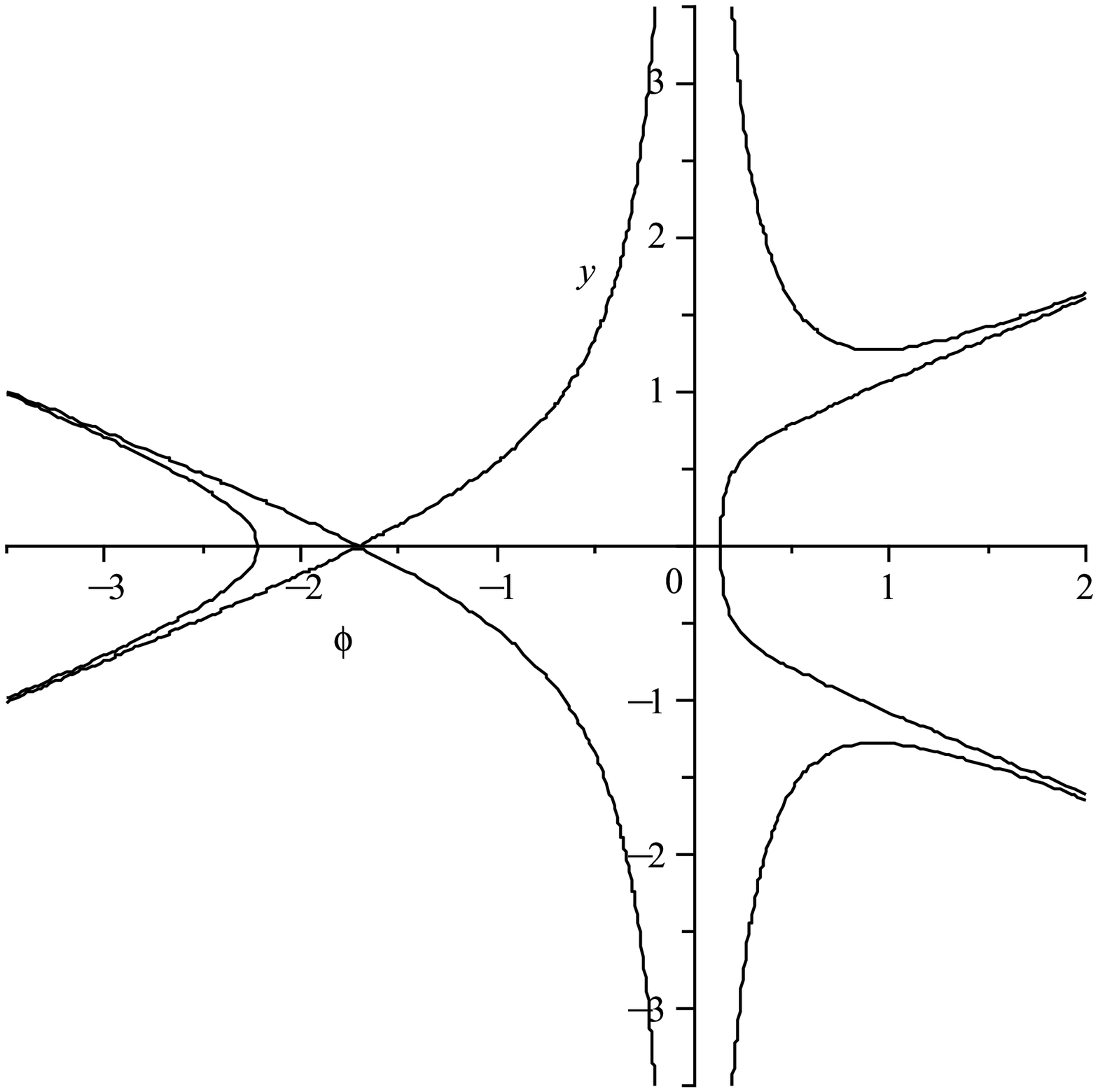}}\hspace{0.01\textwidth}
\subfloat[ ]{ \label{fig:6}
\includegraphics[height=1.in,width=1.2in]{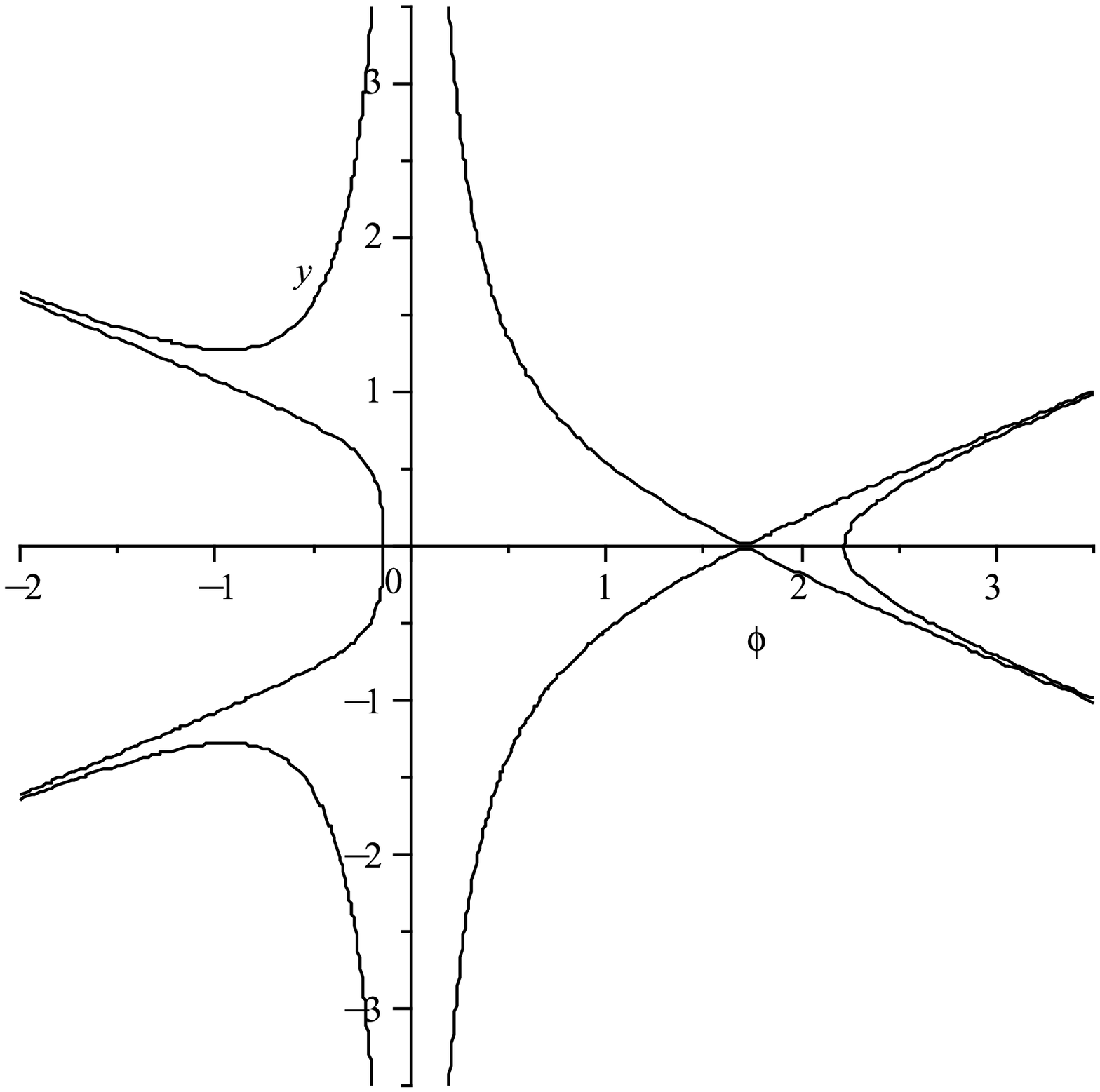}}\hspace{0.01\textwidth}
\subfloat[]{ \label{fig:7}
\includegraphics[height=1in,width=1.2in]{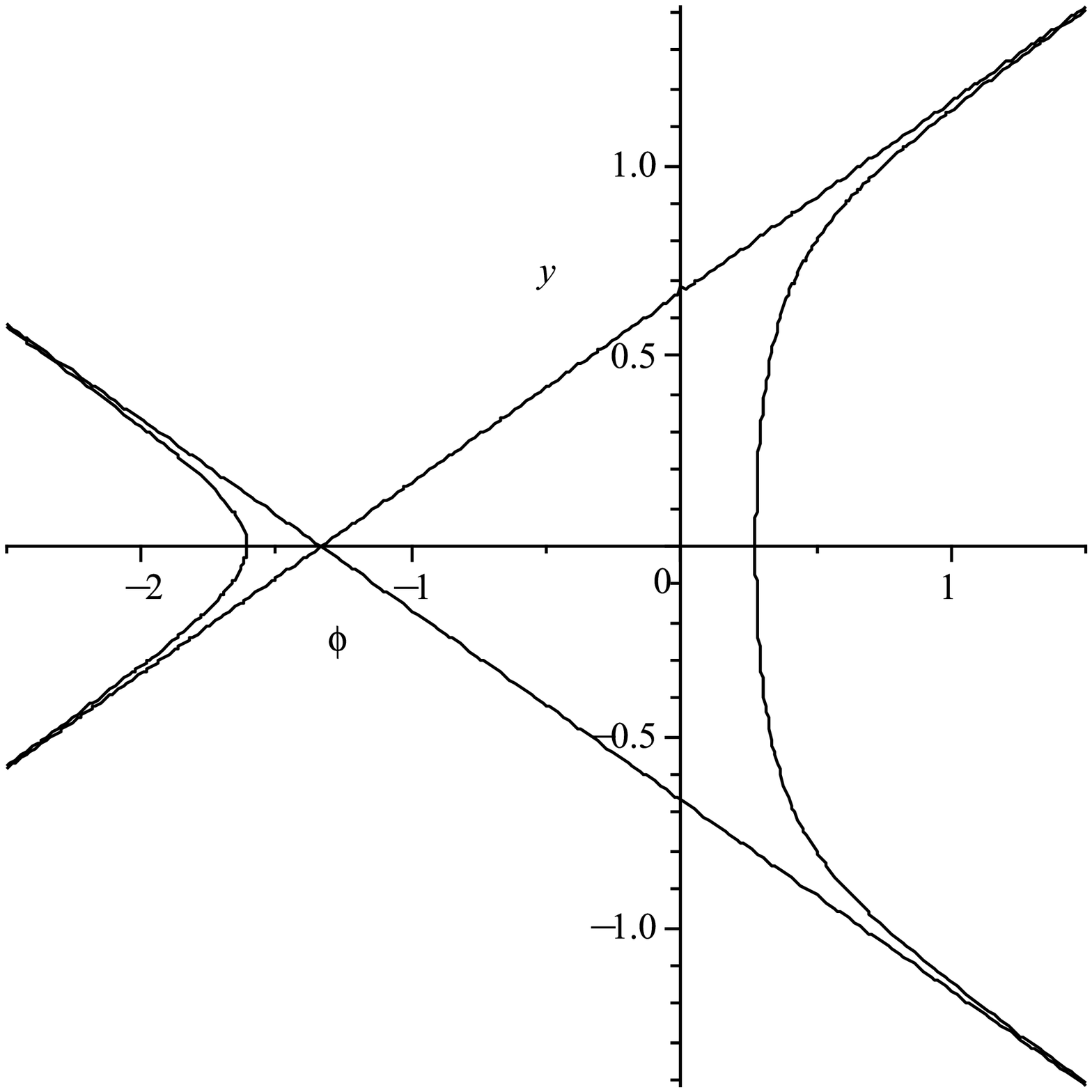}}\hspace{0.01\textwidth}
\subfloat[ ]{ \label{fig:8}
\includegraphics[height=1in,width=1.2in]{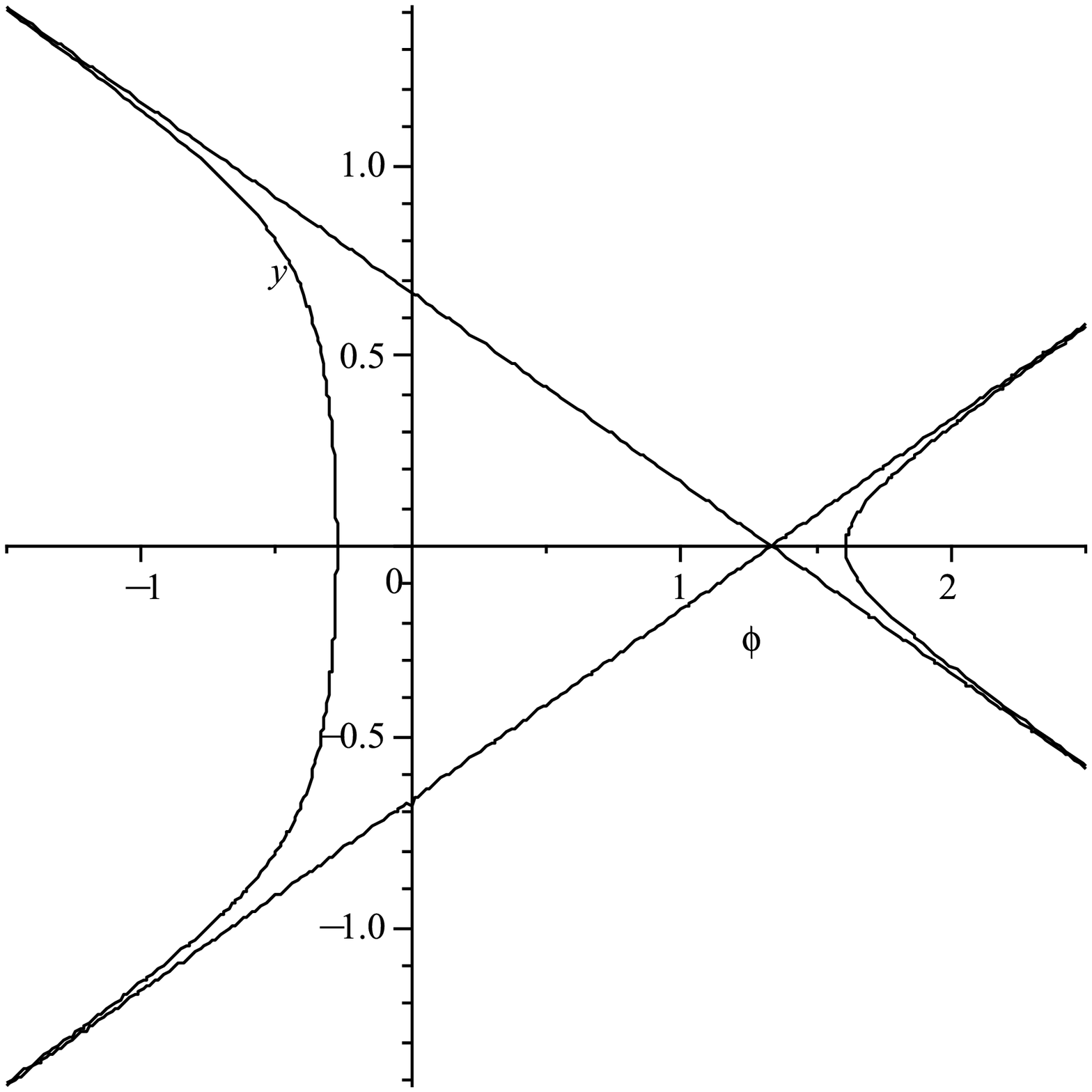}}\\
\subfloat[]{ \label{fig:9}
\includegraphics[height=1in,width=1.2in]{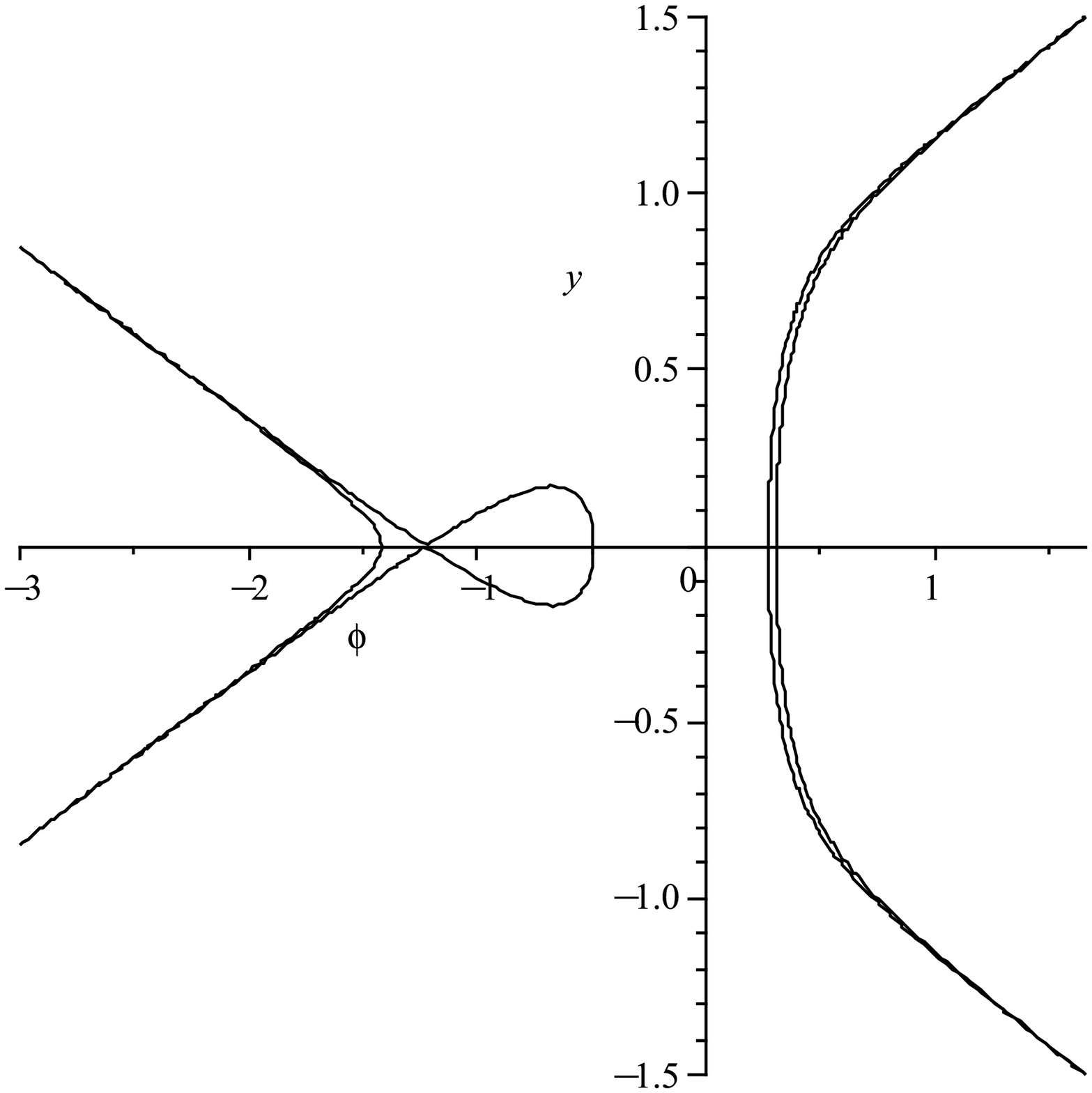}}\hspace{0.01\textwidth}
\subfloat[ ]{ \label{fig:10}
\includegraphics[height=1in,width=1.2in]{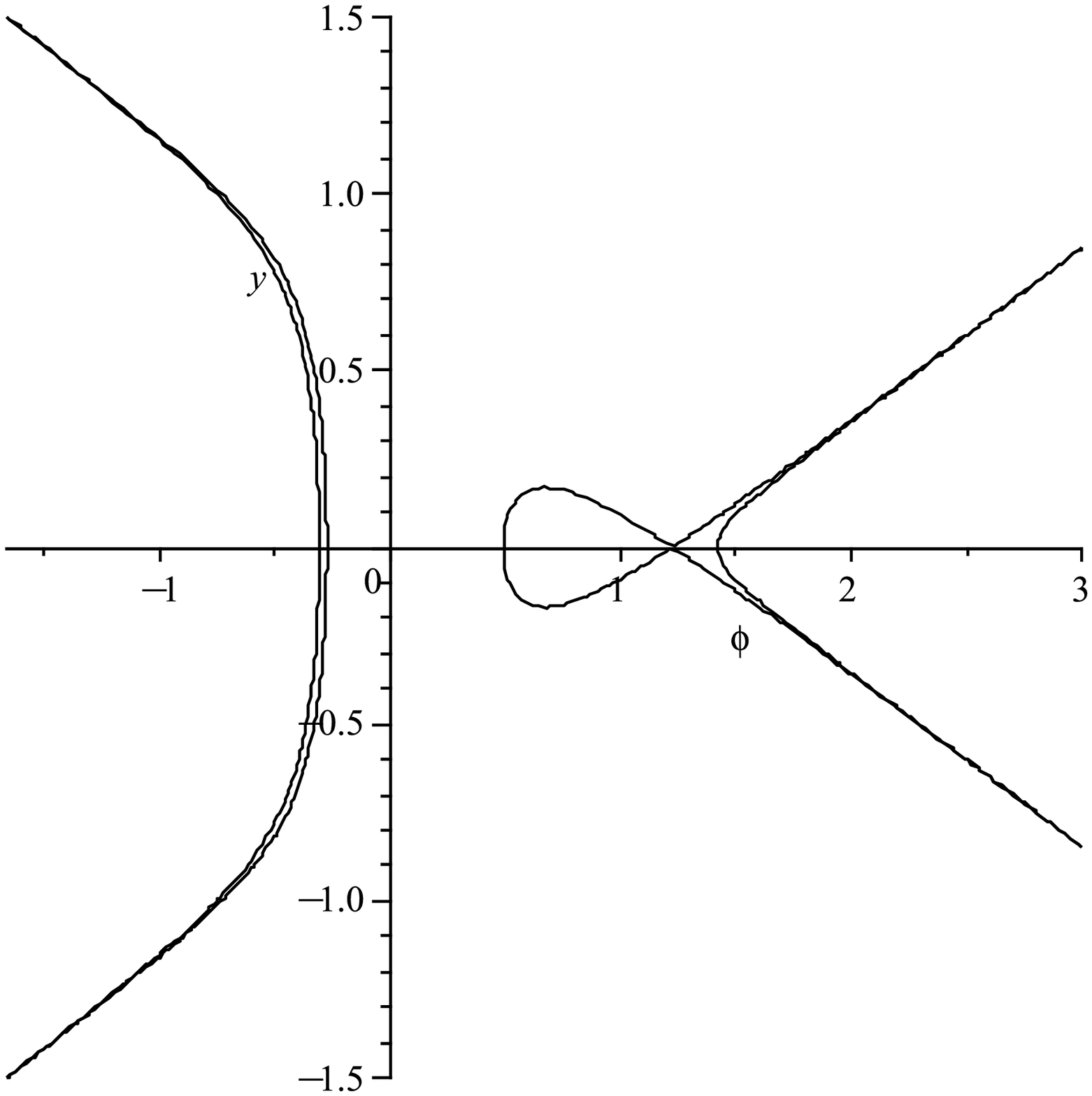}}\hspace{0.01\textwidth}
\subfloat[]{ \label{fig:11}
\includegraphics[height=1in,width=1.2in]{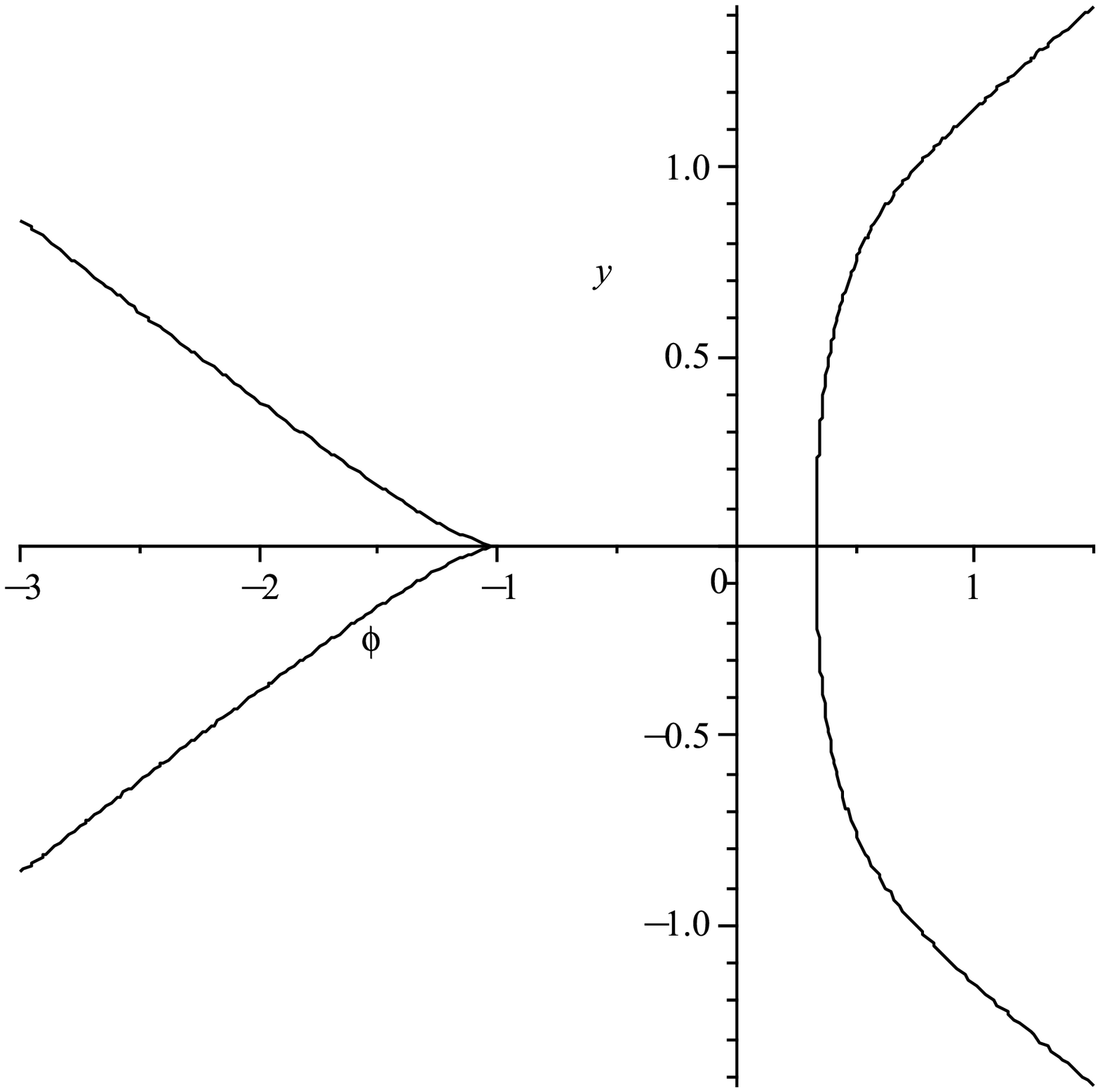}}\hspace{0.01\textwidth}
\subfloat[ ]{ \label{fig:12}
\includegraphics[height=1in,width=1.2in]{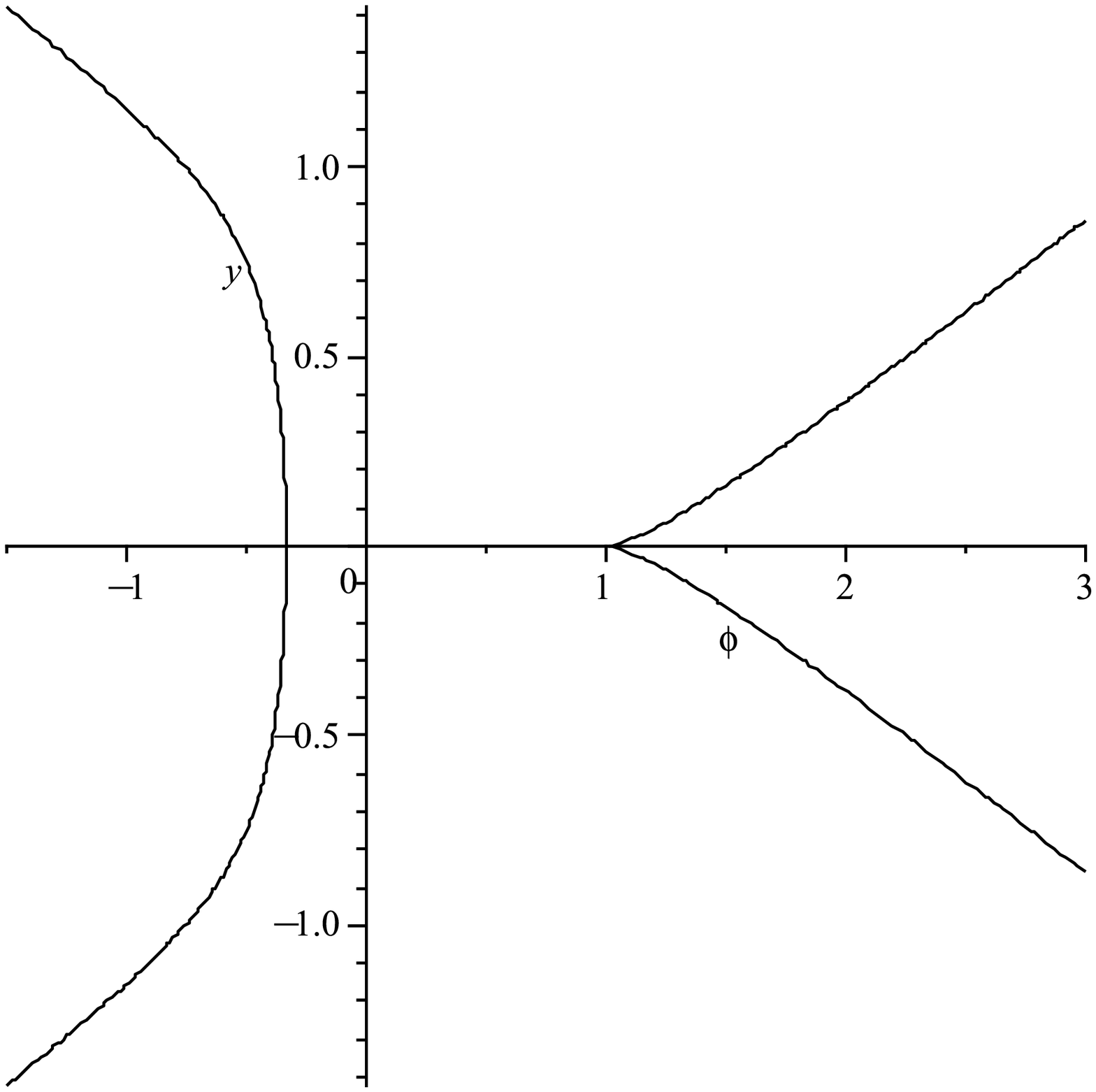}}\\

\caption{Phase portrait of  system (\ref{eq2.6}). (a) $g>0$, $c<0$;
 (b) $g>0$, $c>0$; (c) $g = 0$, $c<0$; (d) $g = 0$, $c>0$; (e) $-\frac{2c^2}{9}<g<0$, $c<0$; (f) $-\frac{2c^2}{9}<g<0$, $c>0$;
 (g) $g=-\frac{2c^2}{9}$, $c<0$; (h) $g=-\frac{2c^2}{9}$, $c>0$; (i) $-\frac{c^2}{4}<g<-\frac{2c^2}{9}$, $c<0$; (j) $-\frac{c^2}{4}<g<-\frac{2c^2}{9}$, $c>0$;
 (k) $g=-\frac{c^2}{4}$, $c<0$; (l) $g=-\frac{c^2}{4}$, $c>0$.}
\end{figure}

\section{Main results}

 \setcounter {equation}{0}

 We state our main result as follows.

\begin{theorem}
\label{th3.1} For given constant $c\neq 0$, let
\begin{equation}
\label{eq3.1}\xi = x - ct,
\end{equation}
\begin{equation}
\label{eq3.2} \varphi _0^\pm =\frac { c  \pm \sqrt {c^2 +4g}}{2},
\end{equation}

(1) When $g>0$ and $c<0$, Eq.(\ref{eq1.7}) has two kink-like wave
solutions $u = \varphi _1 (\xi )$ and $u = \varphi _3 (\xi )$ and
two antikink-like wave solutions $u = \varphi _2 (\xi )$ and $u =
\varphi _4 (\xi )$.
\begin{equation}
\label{eq3.3} \beta _1(\varphi _1) = \beta _1 (a) \exp( -
\frac{1}{2}\xi),\\
\quad \xi \in ( - \infty ,\xi _0^1 ),
\end{equation}
\begin{equation}
\label{eq3.4} \beta _1(\varphi _2) = \beta _1 (a) \exp(
\frac{1}{2}\xi),\\ \quad  \quad \xi \in ( - \xi _0^1 ,\infty ),
\end{equation}
\begin{equation}
\label{eq3.5} \beta _2( \varphi _3) = \beta _2 (b)\exp( -
\frac{1}{2}\xi),\\ \quad \quad  \xi \in ( - \xi _0^3 ,\infty ),
\end{equation}
\begin{equation}
\label{eq3.6} \beta _2( \varphi _4) = \beta _2 (b)\exp(
\frac{1}{2}\xi),\\ \quad \quad  \xi \in ( -\infty ,\xi _0^3  ),
\end{equation}

(2) When $g>0$ and $c>0$, Eq.(\ref{eq1.7}) has two kink-like wave
solutions $u = \varphi _1 (\xi )$ and $u = \varphi _3 (\xi )$ and
two antikink-like wave solutions $u = \varphi _2 (\xi )$ and $u =
\varphi _4 (\xi )$.
\begin{equation}
\label{eq3.7} \beta _1(\varphi _1) = \beta _1 (a) \exp( -
\frac{1}{2}\xi),\\
\quad \xi \in ( - \infty ,\xi _0^5 ),
\end{equation}
\begin{equation}
\label{eq3.8} \beta _1(\varphi _2) = \beta _1 (a) \exp(
\frac{1}{2}\xi),\\ \quad  \quad \xi \in ( - \xi _0^5 ,\infty ),
\end{equation}
\begin{equation}
\label{eq3.9} \beta _2( \varphi _3) = \beta _2 (b)\exp( -
\frac{1}{2}\xi),\\ \quad \quad  \xi \in ( - \xi _0^7 ,\infty ),
\end{equation}
\begin{equation}
\label{eq3.10} \beta _2( \varphi _4) = \beta _2 (b)\exp(
\frac{1}{2}\xi),\\ \quad \quad  \xi \in ( -\infty ,\xi _0^7 ),
\end{equation}

(3) When $-\frac{2c^2}{9}<g\leq0$ and $c<0$, Eq.(\ref{eq1.7}) has a
kink-like wave solution $u = \varphi _5 (\xi )$  and an
antikink-like wave solution $u = \varphi _6 (\xi )$.
\begin{equation}
\label{eq3.11} \beta _1(\varphi _5) = \beta _1 (d) \exp( -
\frac{1}{2}\xi),\\
\quad \xi \in ( - \infty ,\xi _0^9 ),
\end{equation}
\begin{equation}
\label{eq3.12} \beta _1(\varphi _6) = \beta _1 (d) \exp(
\frac{1}{2}\xi),\\ \quad  \quad \xi \in ( - \xi _0^9 ,\infty ),
\end{equation}

(4) When $-\frac{2c^2}{9}<g\leq0$ and $c>0$,  Eq.(\ref{eq1.7}) has a
kink-like wave solution $u = \varphi _7 (\xi )$ and an antikink-like
wave solution $u = \varphi _8 (\xi )$.
\begin{equation}
\label{eq3.13} \beta _2(\varphi _7) = \beta _2 (k) \exp( -
\frac{1}{2}\xi),\\
\quad \xi \in ( - \infty ,\xi _0^{11} ),
\end{equation}
\begin{equation}
\label{eq3.14} \beta _2(\varphi _8) = \beta _2 (k) \exp(
\frac{1}{2}\xi),\\ \quad  \quad \xi \in ( - \xi _0^{11} ,\infty ),
\end{equation}

\noindent where
\begin{equation}
\label{eq3.15} \varphi_1^* = \frac{c-\sqrt {c^2 +3g} }{2},
\end{equation}
\begin{equation}
\label{eq3.16} \varphi_2^* = \frac{c+\sqrt {c^2 +3g} }{2},
\end{equation}
\begin{equation}
\label{eq3.17} l_1 = -\frac{1}{3}(c+3\sqrt {c^2 +4g}),
\end{equation}
\begin{equation}
\label{eq3.18} l_2 = \frac{1}{6}(c^2+6g-c\sqrt {c^2 +4g}),
\end{equation}
\begin{equation}
\label{eq3.19} l_3 =  \frac{2}{3}(c^2+6g-c\sqrt {c^2 +4g}),
\end{equation}
\begin{equation}
\label{eq3.20} m_1 = -\frac{1}{3}(c-3\sqrt {c^2 +4g}),
\end{equation}
\begin{equation}
\label{eq3.21} m_2 = \frac{1}{6}(c^2+6g+c\sqrt {c^2 +4g}),
\end{equation}
\begin{equation}
\label{eq3.22} m_3 =  \frac{2}{3}(c^2+6g+c\sqrt {c^2 +4g}),
\end{equation}
\begin{equation}
\label{eq3.23} a_1 = c^2+4g-c\sqrt {c^2 +4g},
\end{equation}
\begin{equation}
\label{eq3.24} a_2 =c^2+4g+c\sqrt {c^2 +4g},
\end{equation}
\begin{equation}
\label{eq3.25} b_1 =\frac{1}{3}(2c-6\sqrt {c^2 +4g}),
\end{equation}
\begin{equation}
\label{eq3.26} b_2 =\frac{1}{3}(2c+6\sqrt {c^2 +4g}),
\end{equation}
\begin{equation}
\label{eq3.27} \alpha _1 = \frac{c-\sqrt {c^2 +4g}}{2\sqrt {c^2
+4g-c\sqrt {c^2 +4g}}},
\end{equation}
\begin{equation}
\label{eq3.28} \alpha _2 = \frac{c+\sqrt {c^2 +4g}}{2\sqrt {c^2
+4g+c\sqrt {c^2 +4g}}},
\end{equation}
\begin{equation}
\label{eq3.29} \beta _1(\varphi) =  \frac{(2\sqrt {\varphi^2 + l_1
\varphi + l_2 } + 2\varphi + l_1 )(\varphi - \varphi _0^ - )^{\alpha
_1 }}{(2\sqrt {a_1 } \sqrt {\varphi^2 + l_1 \varphi + l_2 } + b_1
\varphi + l_3 )^{\alpha _1 }},
\end{equation}
\begin{equation}
\label{eq3.30} \beta _2(\varphi) = \frac{(2\sqrt {\varphi^2 + m_1
\varphi + m_2 } + 2\varphi + m_1 )(\varphi - \varphi _0^ + )^{\alpha
_2 }}{(2\sqrt {a_2 } \sqrt {\varphi^2 + m_1 \varphi + m_2 } + b_2
\varphi + m_3 )^{\alpha _2 }},
\end{equation}
\begin{equation}
\label{eq3.31} \xi _0^1 = 2\ln (\beta _1(a) / \beta _1(\frac{c}{2}
)),
\end{equation}
\begin{equation}
\label{eq3.32} \xi _0^3 = 2\ln (\beta _2(\varphi_2^*) / \beta _2(b)
)
\end{equation}
\begin{equation}
\label{eq3.33} \xi _0^5 = 2\ln (\beta _1(a) / \beta _1(\varphi_1^*)
),
\end{equation}
\begin{equation}
\label{eq3.34} \xi _0^7 = 2\ln (\beta _2(\frac{c}{2}) / \beta _2(b)
),
\end{equation}
\begin{equation}
\label{eq3.35} \xi _0^9 = 2\ln (\beta _1(d) / \beta _1(\frac{c}{2})
),
\end{equation}
\begin{equation}
\label{eq3.36} \xi _0^{11} = 2\ln (\beta _2(\frac{c}{2}) / \beta
_2(k) ),
\end{equation}

\noindent $a$, $b$, $d$, $k$ are four constants satisfying $\varphi
_1 (0) = \varphi _2 (0) = a$, $\varphi _3 (0) = \varphi _4 (0) = b$,
$\varphi _5 (0) = \varphi _6 (0) = d$, $\varphi _7 (0) = \varphi _8
(0) = k$, and there are inequalities $\varphi _0^ - < a <
\frac{c}{2} <0<\varphi_2^* < b < \varphi _0^ + $ for $c<0$, $\varphi
_0^ - < a <\varphi_1^*<0 <\frac{c}{2}  < b < \varphi _0^ + $ for
$c>0$, $\varphi _0^ - < d< \frac{c}{2} < \varphi _0^ + <0 $ for
$c<0$ and $ 0<\varphi _0^ - < \frac{c}{2} <k < \varphi _0^ + $ for
$c>0$.
 \end{theorem}

We will give the proof of this theorem in Section 4. Now we take a
set of data and employ Maple to display the graphs of $u = \varphi
_i (\xi )(i = 1,2,3,4,5,6,7,8)$.

\begin{example}
\label{ex3.1}
 Taking $g=5$, $c=-1$  (corresponding to (1) in Theorem (\ref{th3.1})), it follows that $\varphi
_0^ - = - 2.79129$ , $\varphi _0^ + = 1.79129$, $l_1 = - 4.24924$,
$l_2 = 5.93043$, $l_3 = 23.7217$, $a_1 = 25.5826$, $b_1 = - 9.83182$
, $\alpha _1 = - 0.551865$. Further, choosing $a = -0.75 \in
(\varphi _0^ - ,\frac{c}{2})$, we obtain $\xi _0^1 = 0.0482492$. We
present the graphs of the solutions $\varphi _1 (\xi )$ and $\varphi
_2 (\xi )$ in Fig.2 (a) and (b), respectively. Meanwhile, we get
$m_1 = 4.51591$, $m_2 = 4.4029$, $m_3 =17.6116$, $a_2 = 16.4174$,
$b_2 = 8.49848$, $\alpha _2 = 0.442092$, $\varphi_2^*=1.5$. Further,
choosing $b = 1.6\in (\varphi_2^*,\varphi _0^ + )$, we get $\xi _0^3
= 0.343656$. The graphs of the solutions $\varphi _3 (\xi )$ and
$\varphi _4 (\xi )$ are presented in Fig.2(c) and (d), respentively.
The graphs in Fig.2 show that $\varphi _1 (\xi )$ and $\varphi _3
(\xi )$ are two kink-like wave solutions and $\varphi _2 (\xi )$ and
$\varphi _4 (\xi )$ are two antikink-like wave solutions.
\end{example}

\begin{figure}[h]
\centering \subfloat[]{\label{fig:1}
\includegraphics[height=1.8in,width=2.4in]{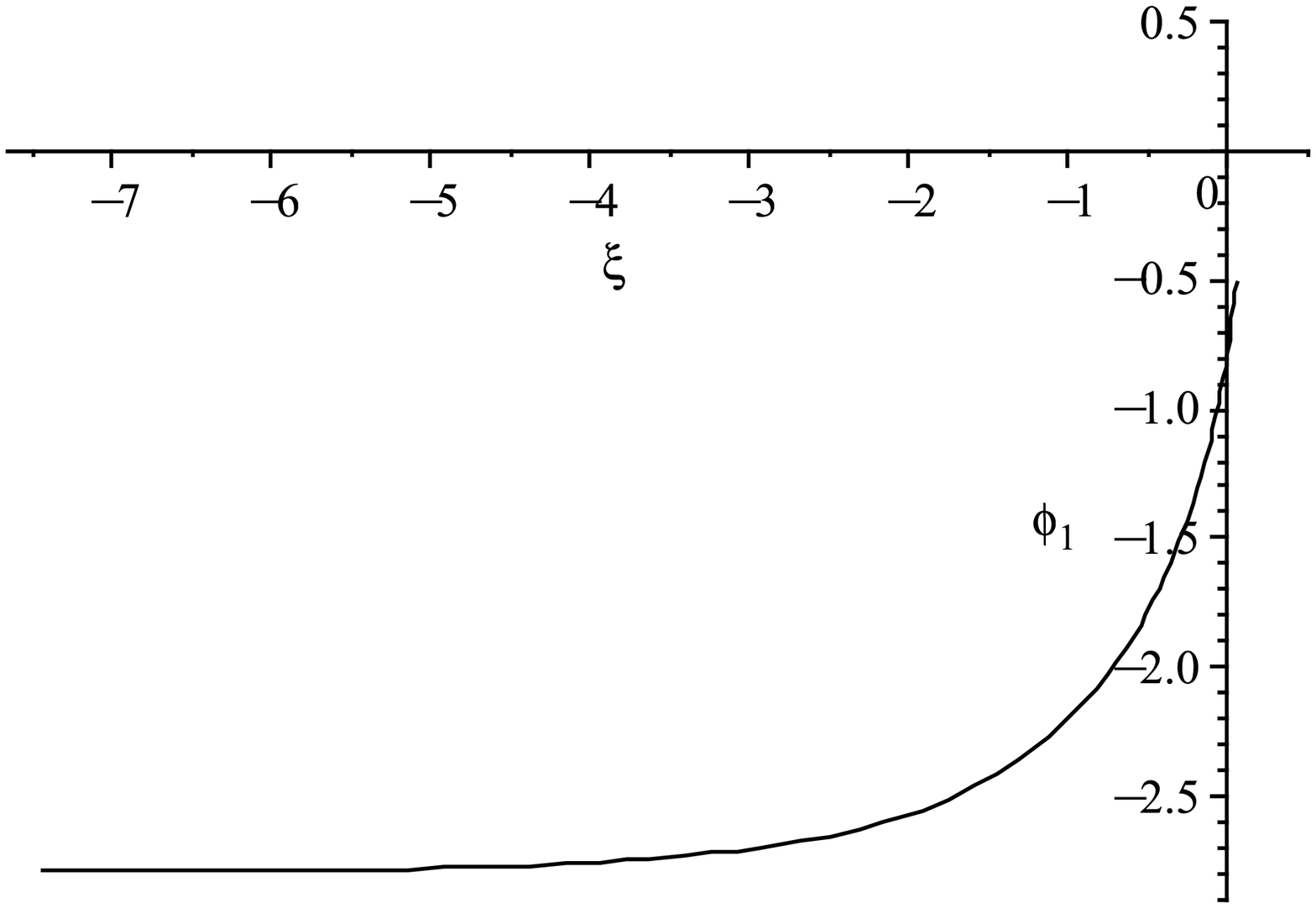}}\hspace{0.03\textwidth}
\subfloat[ ]{ \label{fig:2}
\includegraphics[height=1.8in,width=2.4in]{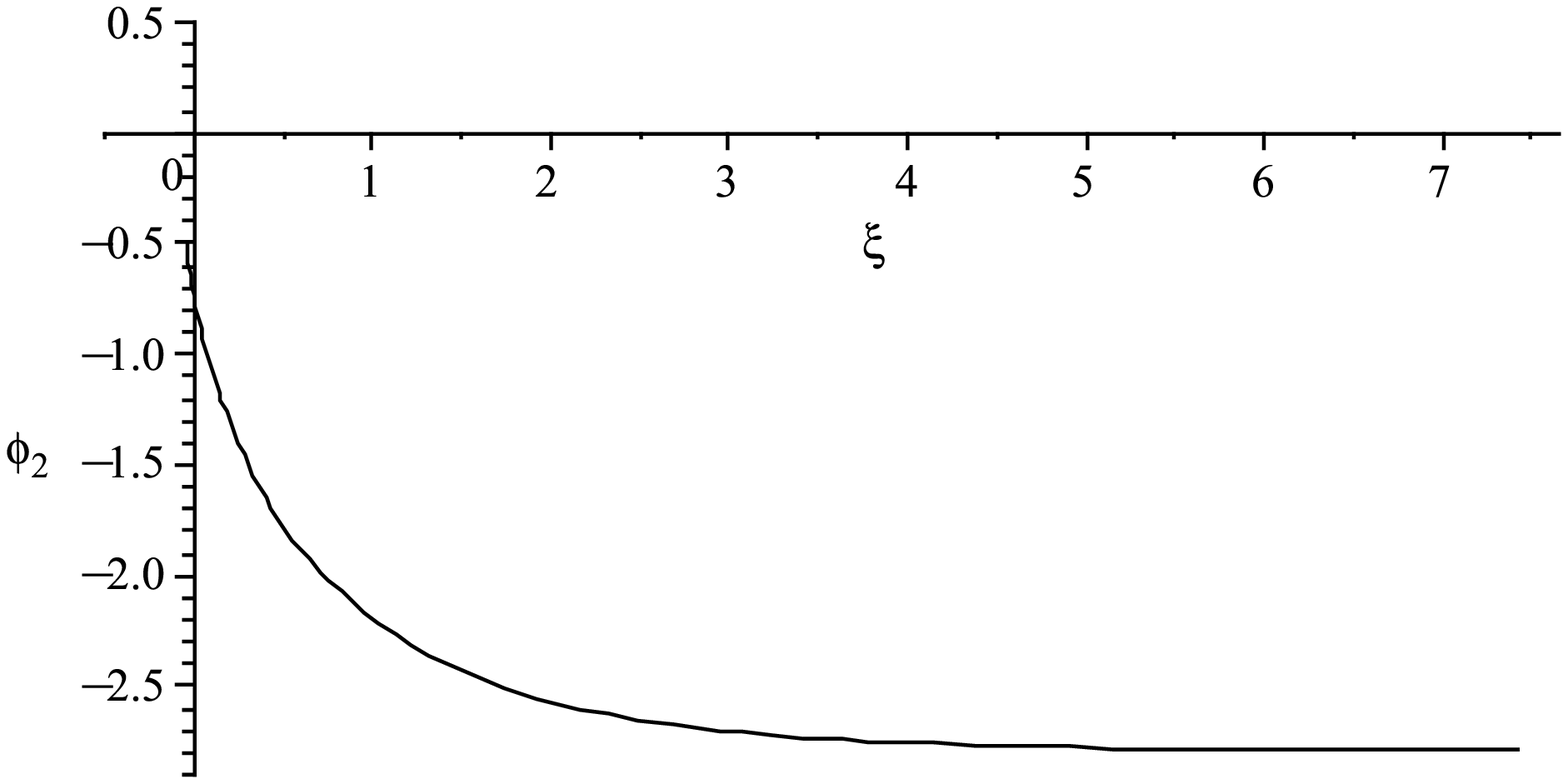}}\\
\subfloat[]{ \label{fig:3}
\includegraphics[height=1.8in,width=2.4in]{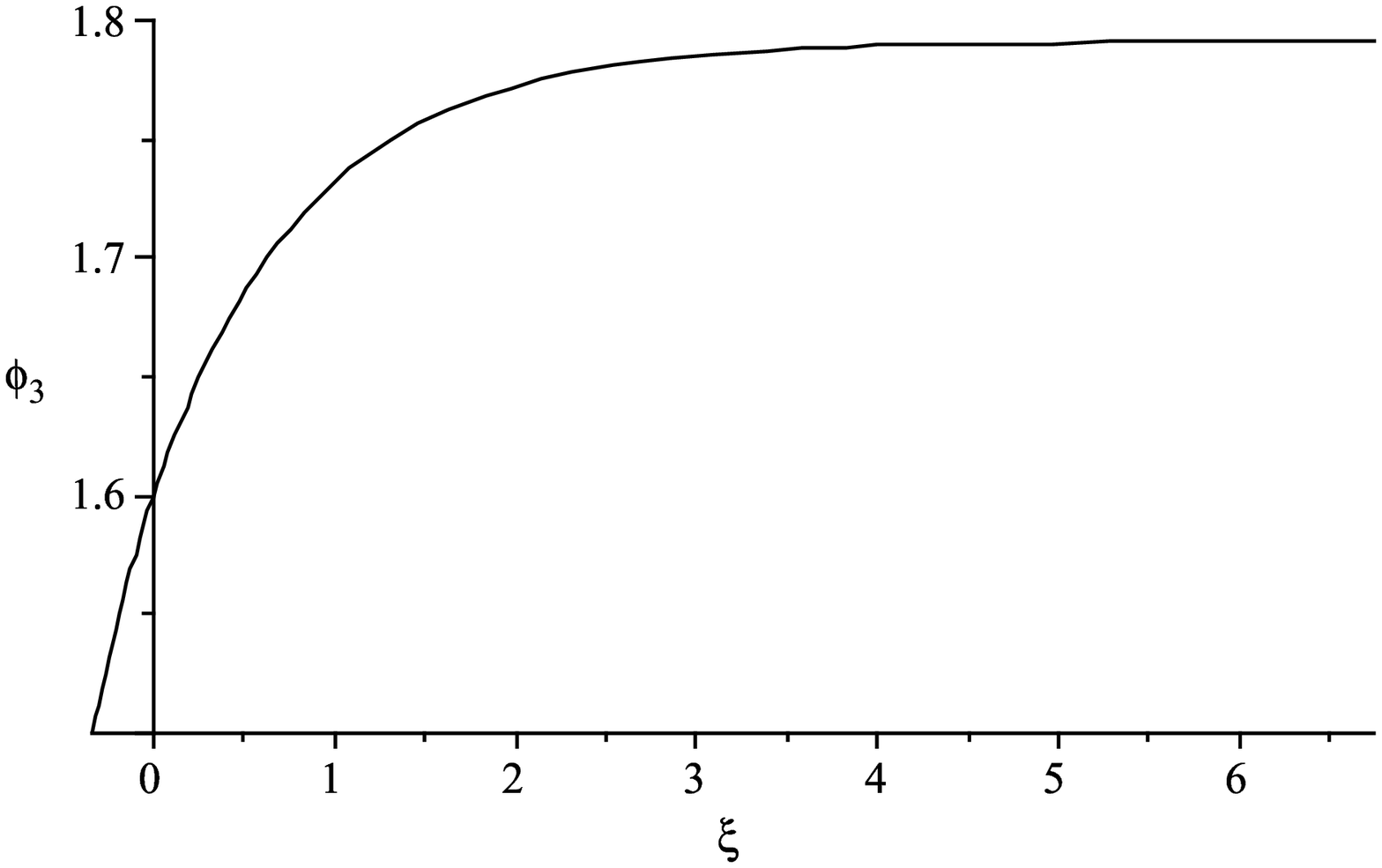}}\hspace{0.03\textwidth}
\subfloat[ ]{ \label{fig:4}
\includegraphics[height=1.79in,width=2.4in]{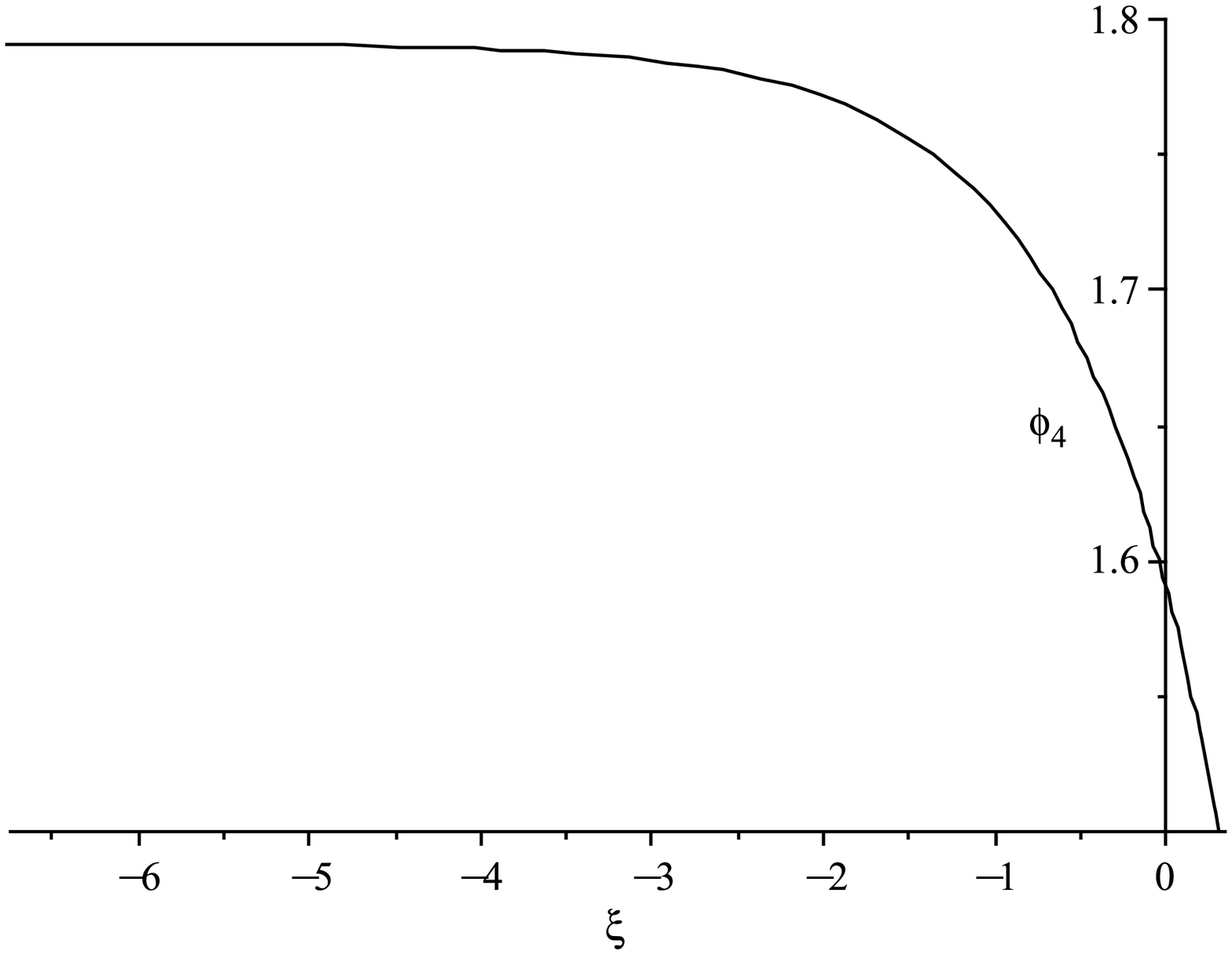}}\\

\caption{The graphs of $\varphi_i(\xi)(i=1,2,3,4)$ when $g=5$,
$c=-1$, $a=-0.75$, $b=1.6$.}
\end{figure}

\begin{example}
\label{ex3.2}
 Taking $g = 5$, $c =1$ (corresponding to (2) in Theorem (\ref{th3.1})), it follows that $\varphi
_0^ - = - 1.79129$ , $\varphi _0^ + = 2.79129$, $l_1 = - 4.51591$,
$l_2 = 4.4029$, $l_3 = 17.6116$, $a_1 = 16.4174$, $b_1 = - 8.49848$
, $\alpha _1 = - 0.442092$, $\varphi_1^*=-1.5$. Further, choosing $a
= -1.6 \in (\varphi _0^ - ,\varphi_1^*)$, we obtain $\xi _0^5 =
0.343656$. We present the graphs of the solutions $\varphi _1 (\xi
)$ and $\varphi _2 (\xi )$ in Fig.3 (a) and (b), respectively.
Meanwhile, we get $m_1 = 4.249241$, $m_2 = 5.93043$, $m_3 =23.7214$,
$a_2 = 25.58264$, $b_2 = 9.93182$, $\alpha _2 = 0.551865$. Further,
choosing $b = 2\in (\frac{c}{2},\varphi _0^ + )$, we get $\xi _0^7 =
0.773847$. The graphs of the solutions $\varphi _3 (\xi )$ and
$\varphi _4 (\xi )$ are presented in Fig.3(c) and (d), respentively.
The graphs in Fig.3 show that $\varphi _1 (\xi )$ and $\varphi _3
(\xi )$ are two kink-like wave solutions and $\varphi _2 (\xi )$ and
$\varphi _4 (\xi )$ are two antikink-like wave solutions.
\end{example}

\begin{figure}[h]
\centering \subfloat[]{\label{fig:1}
\includegraphics[height=1.8in,width=2.4in]{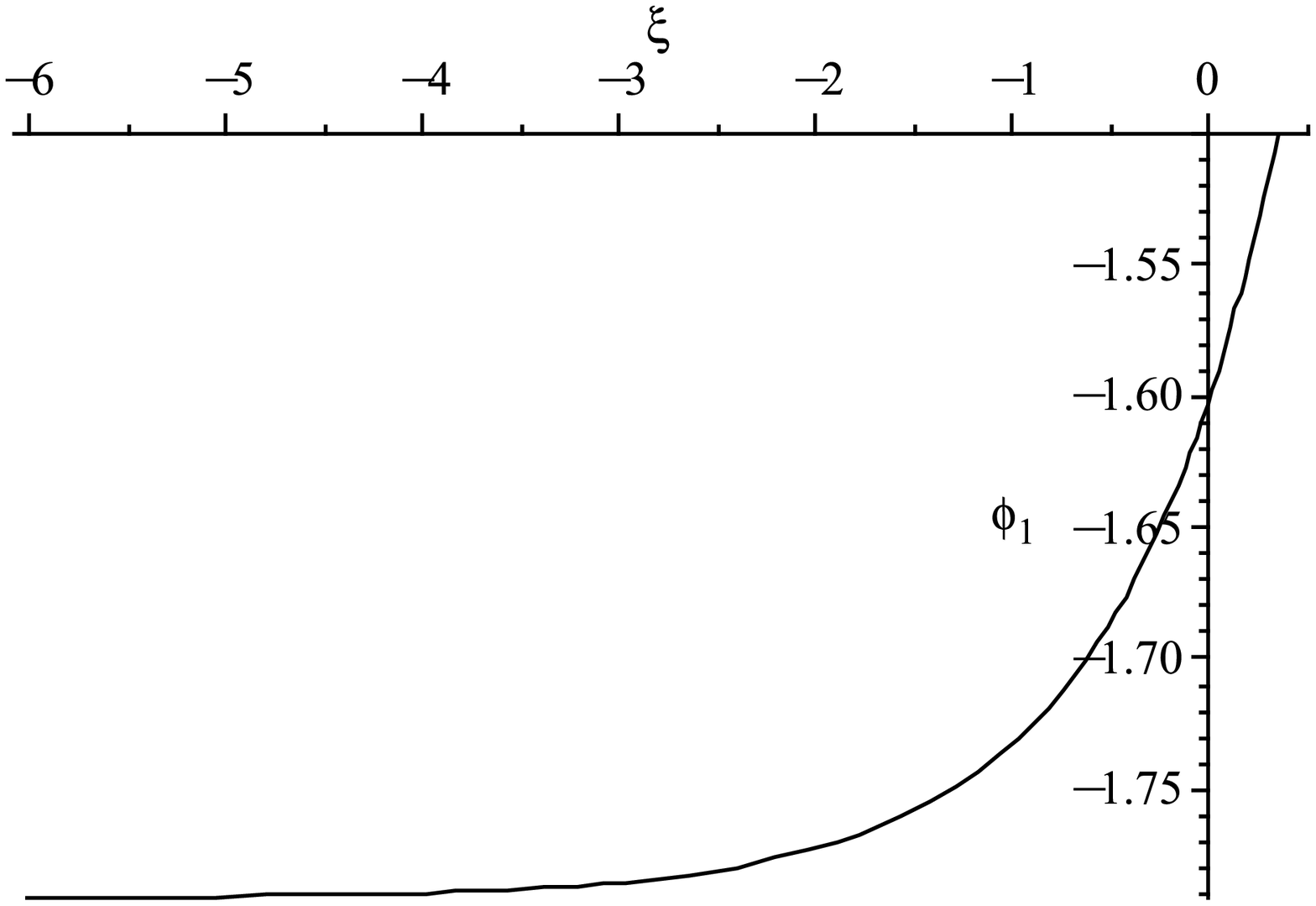}}\hspace{0.03\textwidth}
\subfloat[ ]{ \label{fig:2}
\includegraphics[height=1.8in,width=2.4in]{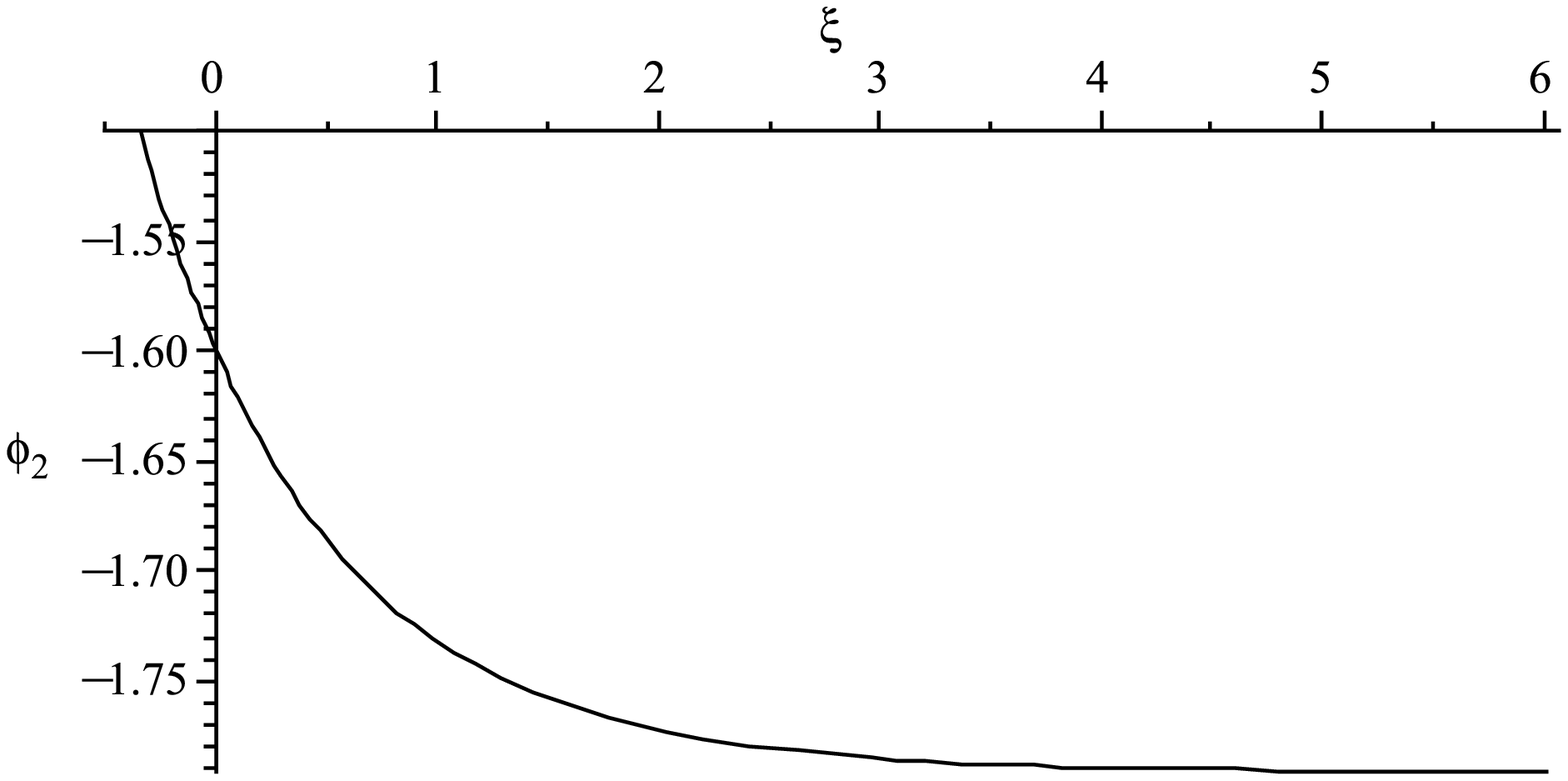}}\\
\subfloat[]{ \label{fig:3}
\includegraphics[height=1.8in,width=2.4in]{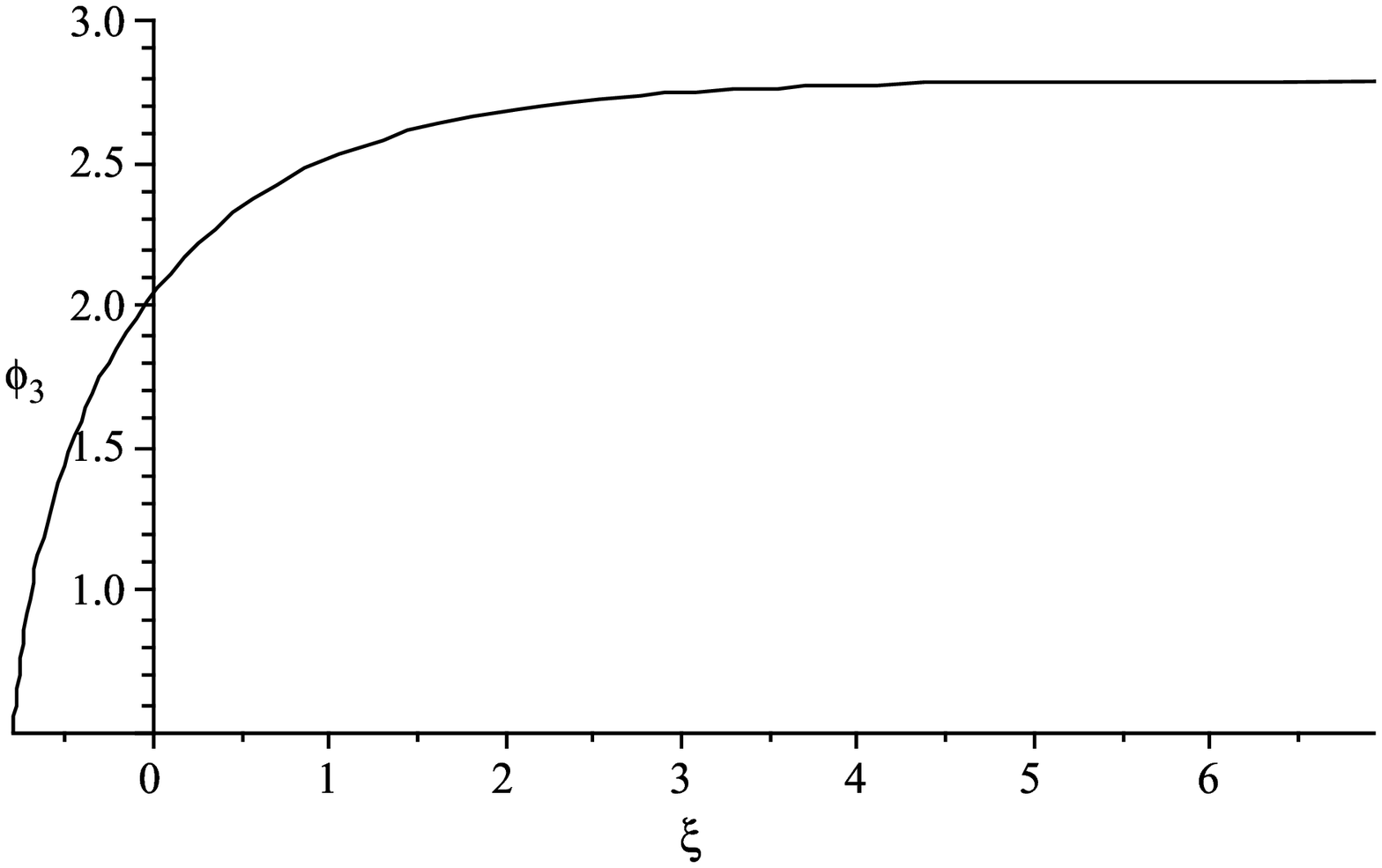}}\hspace{0.03\textwidth}
\subfloat[ ]{ \label{fig:4}
\includegraphics[height=1.75in,width=2.4in]{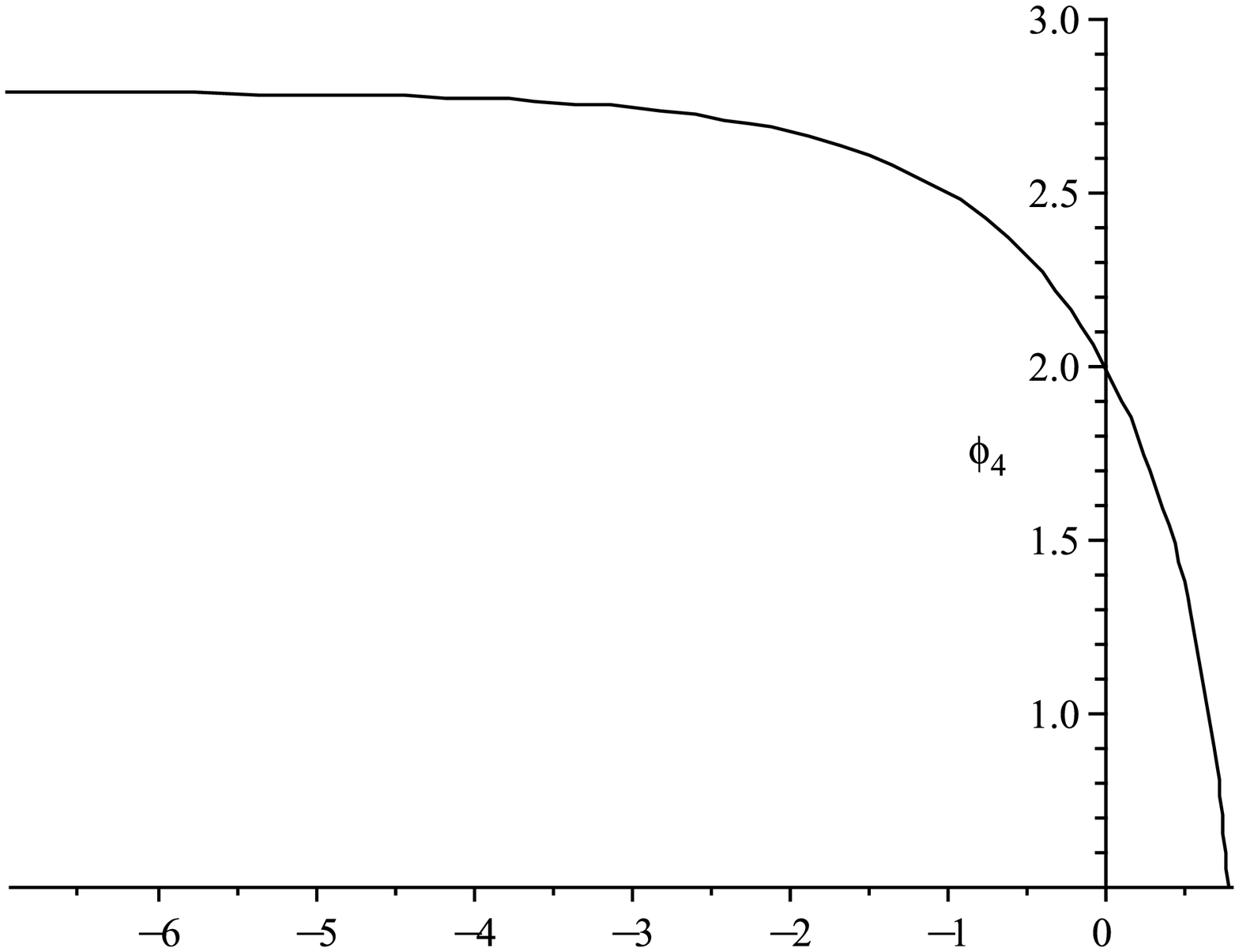}}\\

\caption{The graphs of $\varphi_i(\xi)(i=1,2,3,4)$ when $g=5$,
$c=1$, $a=-1.6$, $b=2$.}
\end{figure}

\begin{example}
\label{ex3.3}
  Taking $g = -0.5$, $c =-2$ (corresponding to (3) in Theorem (\ref{th3.1})), it follows that $\varphi
_0^ - = - 1.70711$ , $\varphi _0^ + = -0.292893$, $l_1 = -
0.747547$, $l_2 = 0.638071$, $l_3 = 2.5528$, $a_1 = 4.82843$, $b_1 =
- 4.16176$ , $\alpha _1 = - 0.776887$. Further, choosing $d = -1.2
\in (\varphi _0^ - ,\frac{c}{2})$, we obtain $\xi _0^9 = 0.448123$.
We present the graphs of the solutions $\varphi _5 (\xi )$ and
$\varphi _6 (\xi )$ in Fig.4 (a) and (b), respectively. The graphs
in Fig.4 show that $\varphi _5 (\xi )$ is a kink-like wave solution
and $\varphi _6 (\xi )$ is an antikink-like wave solution.
\end{example}

\begin{figure}[h]
\centering \subfloat[]{\label{fig:1}
\includegraphics[height=2in,width=2.4in]{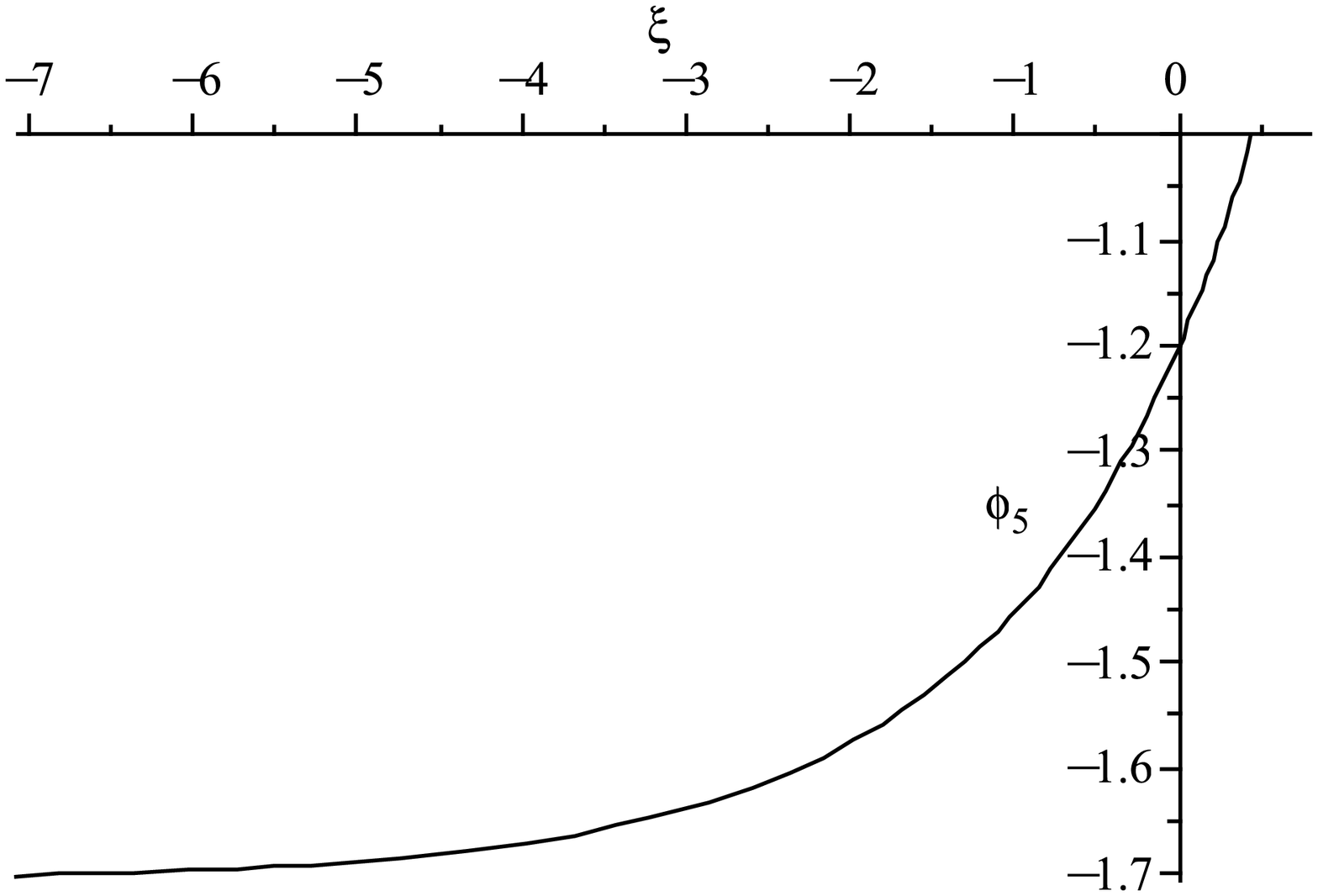}}\hspace{0.04\textwidth}
\subfloat[ ]{ \label{fig:2}
\includegraphics[height=2in,width=2.4in]{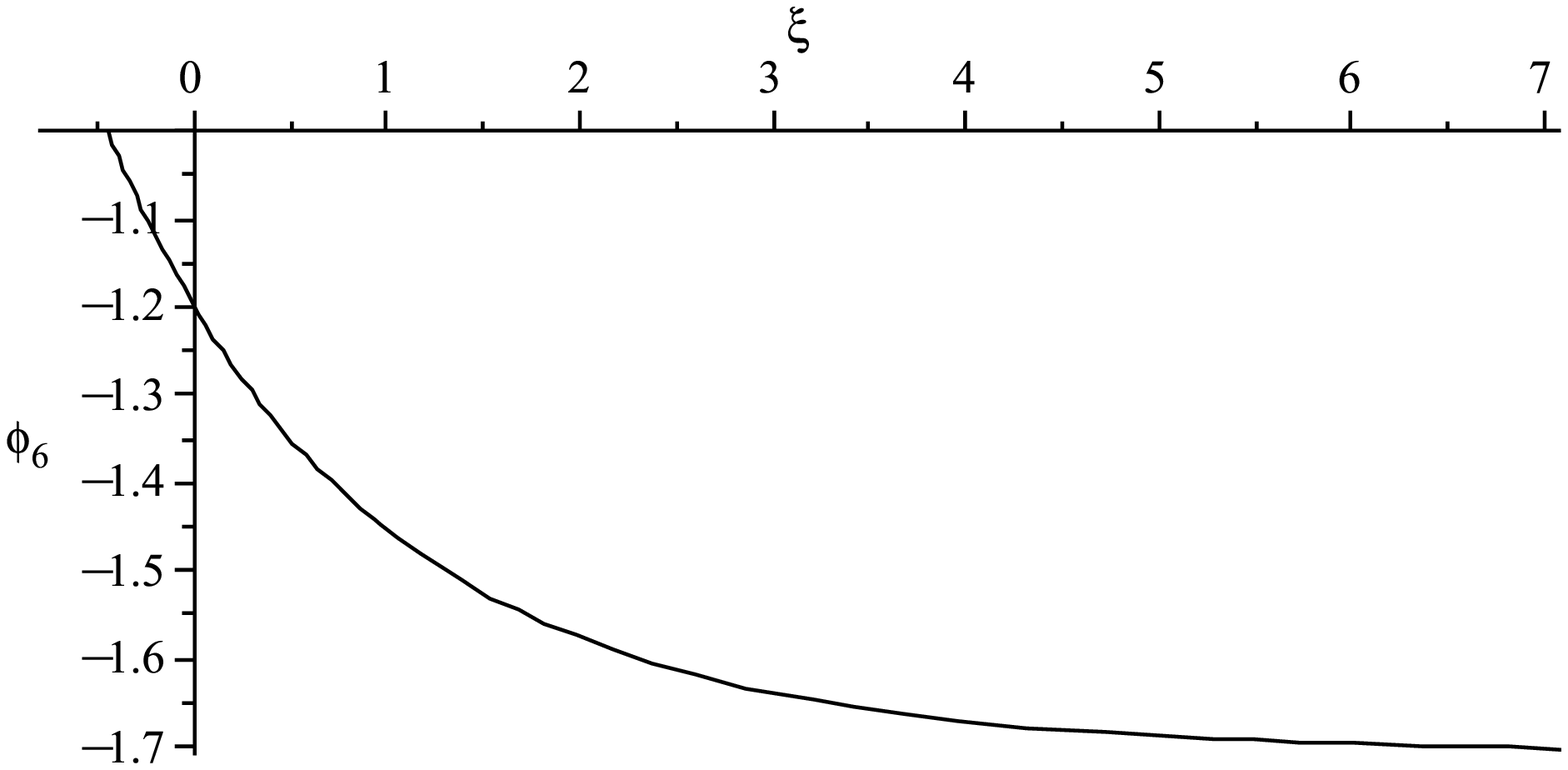}}\\
\caption{The graphs of $\varphi_i(\xi)(i=5,6)$ when $g=-0.5$,
$c=-2$, $d=-1.2$.}
\end{figure}

\begin{example}
\label{ex3.4}
 Taking $g = -0.5$, $c =2$ (corresponding to (4) in Theorem (\ref{th3.1})), it follows that $\varphi
_0^ - = 0.292893$ , $\varphi _0^ + = 1.70711$, $m_1 =0.747547$, $m_2
= 0638071$, $m_3 =2.55228$, $a_2 = 4.82843$, $b_2 = 4.1676$, $\alpha
_2 = 0.776887$. Further, choosing $k = 1.2 \in (\frac{c}{2},\varphi
_0^ + )$, we get $\xi _0^{11} = 0.448123$. The graphs of the
solutions $\varphi _7 (\xi )$ and $\varphi _8 (\xi )$ are presented
in Fig.5 (a) and (b), respentively. The graphs in Fig.5 show that
$\varphi _7 (\xi )$ is a kink-like wave solutions and $\varphi _8
(\xi )$ is an antikink-like wave solutions.
\end{example}

\begin{figure}[h]
\centering \subfloat[]{\label{fig:1}
\includegraphics[height=2in,width=2.4in]{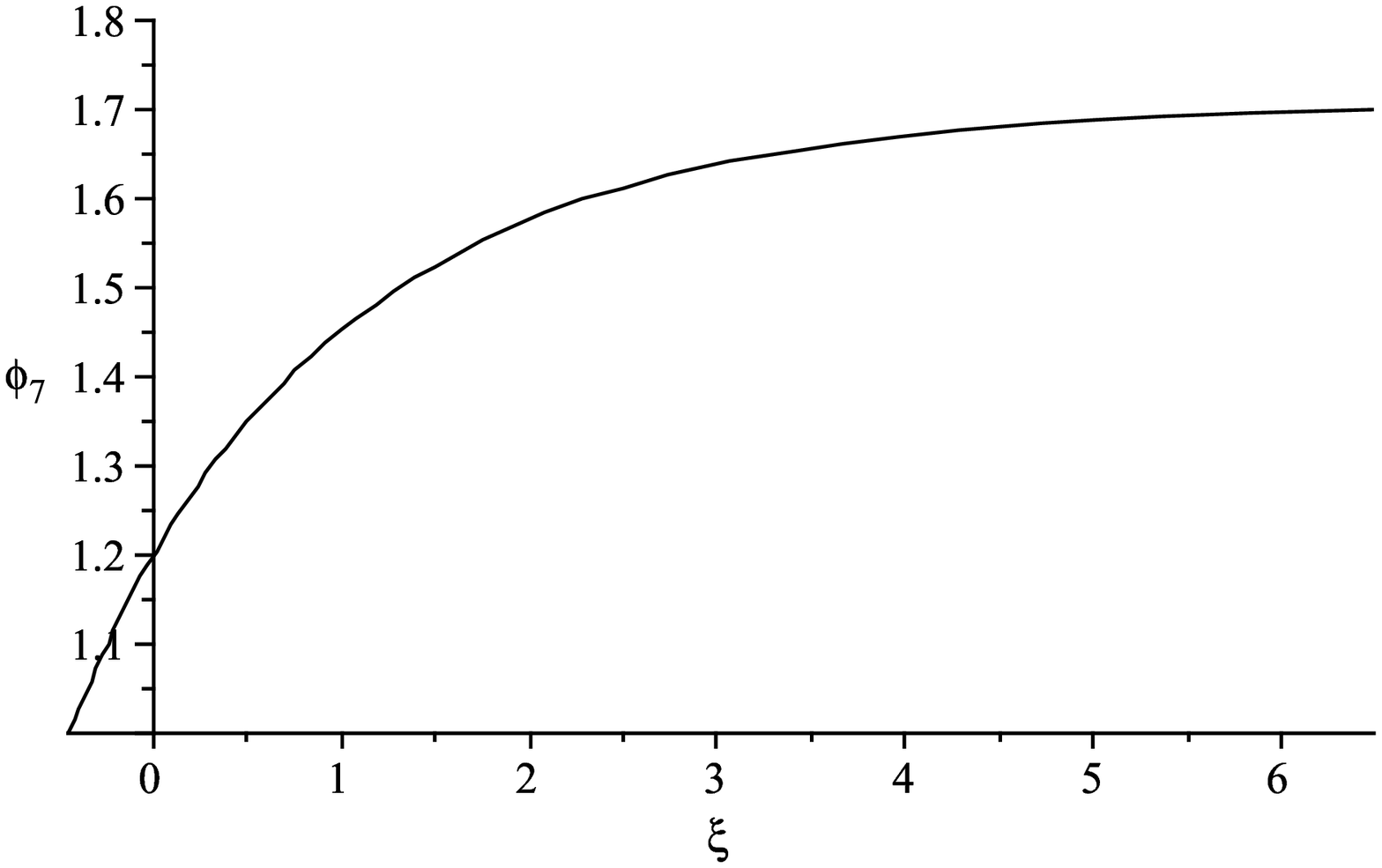}}\hspace{0.04\textwidth}
\subfloat[ ]{ \label{fig:2}
\includegraphics[height=2in,width=2.4in]{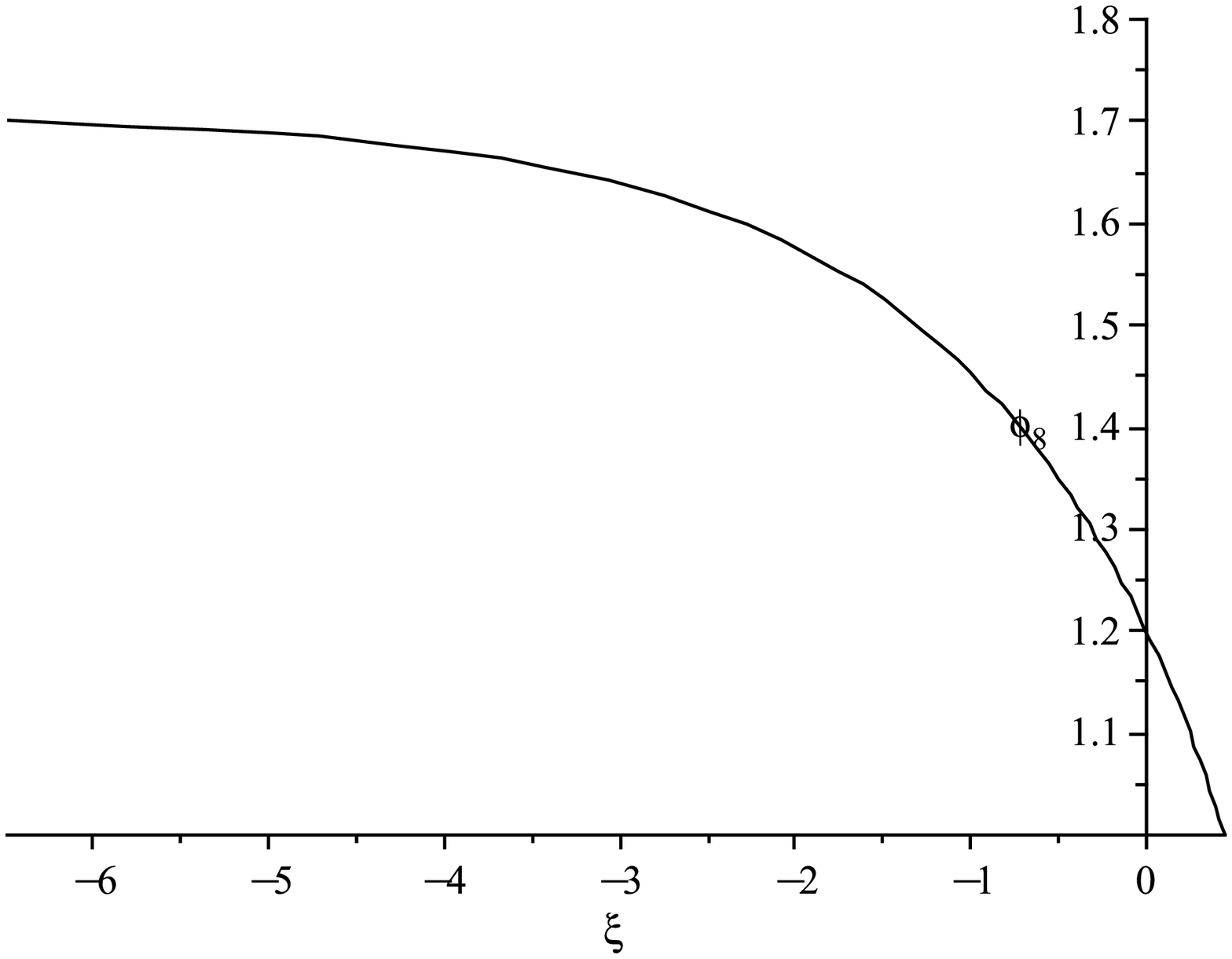}}\\

\caption{The graphs of $\varphi_i(\xi)(i=7,8)$ when $g=-0.5$, $c=2$,
$k=1.2$.}
\end{figure}

\section{ Proof of main results}
\setcounter {equation}{0}

Suppose $g>0$ and $c<0$, then system (\ref{eq2.4}) has two saddle
points $(\varphi _0^ - ,0)$ and $(\varphi _0^ + ,0)$. There are four
orbits connecting with $(\varphi _0^ - ,0)$. We use $l_{\varphi _0^
- }^2 $ to denote the two orbits lying on the right side of
$(\varphi _0^ - ,0)$ (see Fig.6(a)). Meanwhile, there are four
orbits connecting with $(\varphi _0^ + ,0)$. We employ $l_{\varphi
_0^ + }^1 $ and $l_{\varphi _0^ + }^2 $ to denote the two orbits
lying on the left side of $(\varphi _0^ + ,0)$ (see Fig.6(a)).

\begin{figure}[h]
\centering \subfloat[]{\label{fig:1}
\includegraphics[height=1.6in,width=2in]{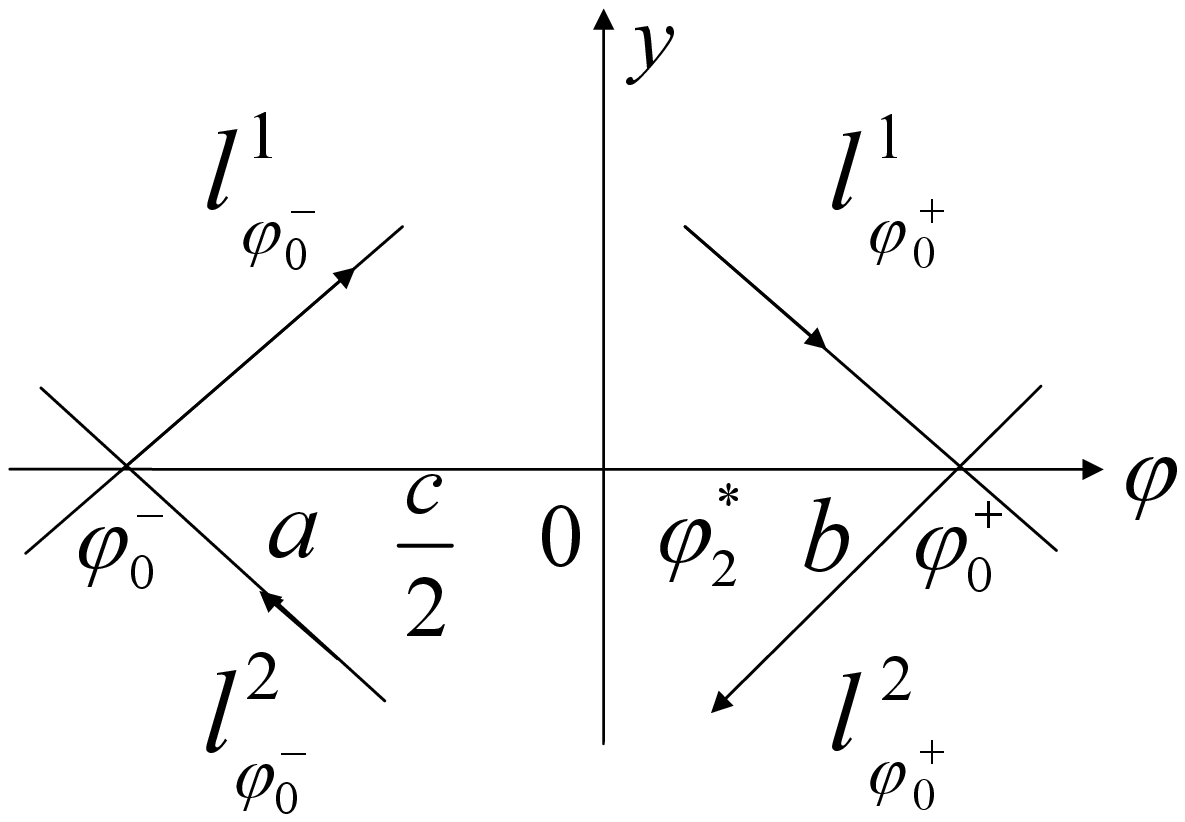}}\hspace{0.06\textwidth}
\subfloat[ ]{ \label{fig:2}
\includegraphics[height=1.6in,width=2in]{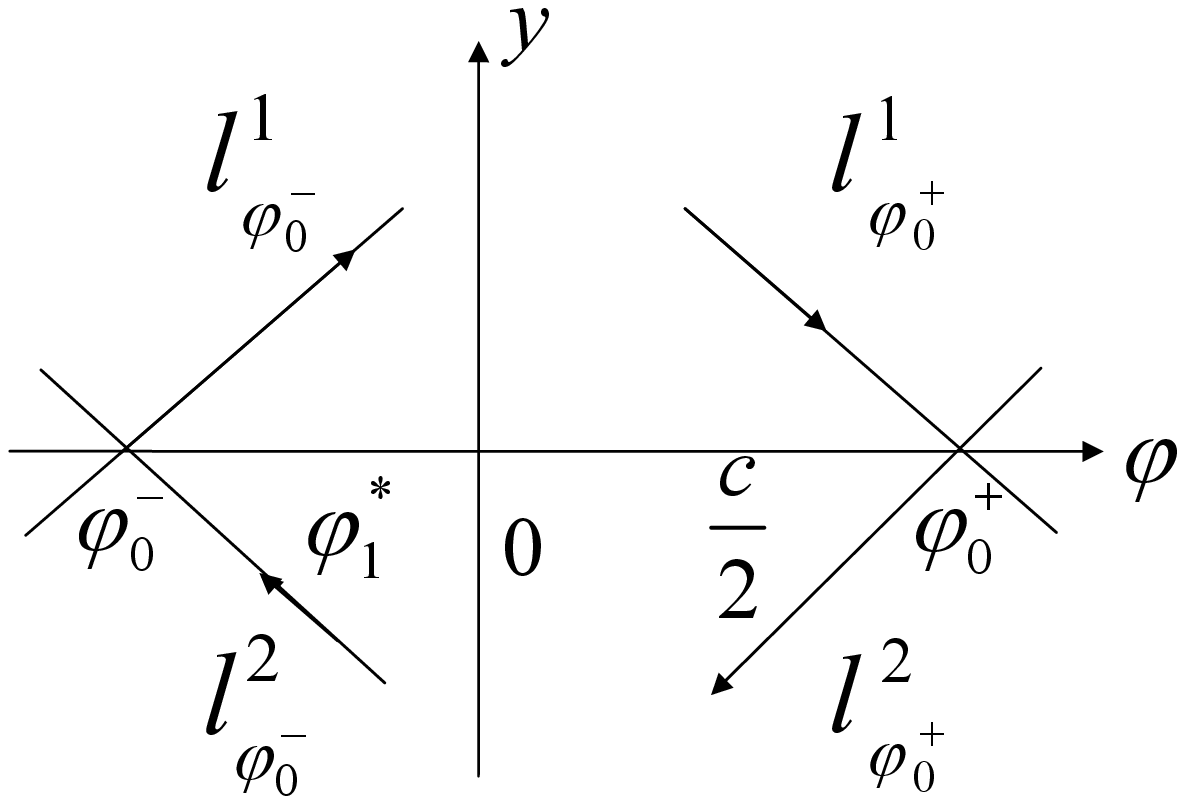}}\\
\subfloat[]{ \label{fig:3}
\includegraphics[height=1.6in,width=1.8in]{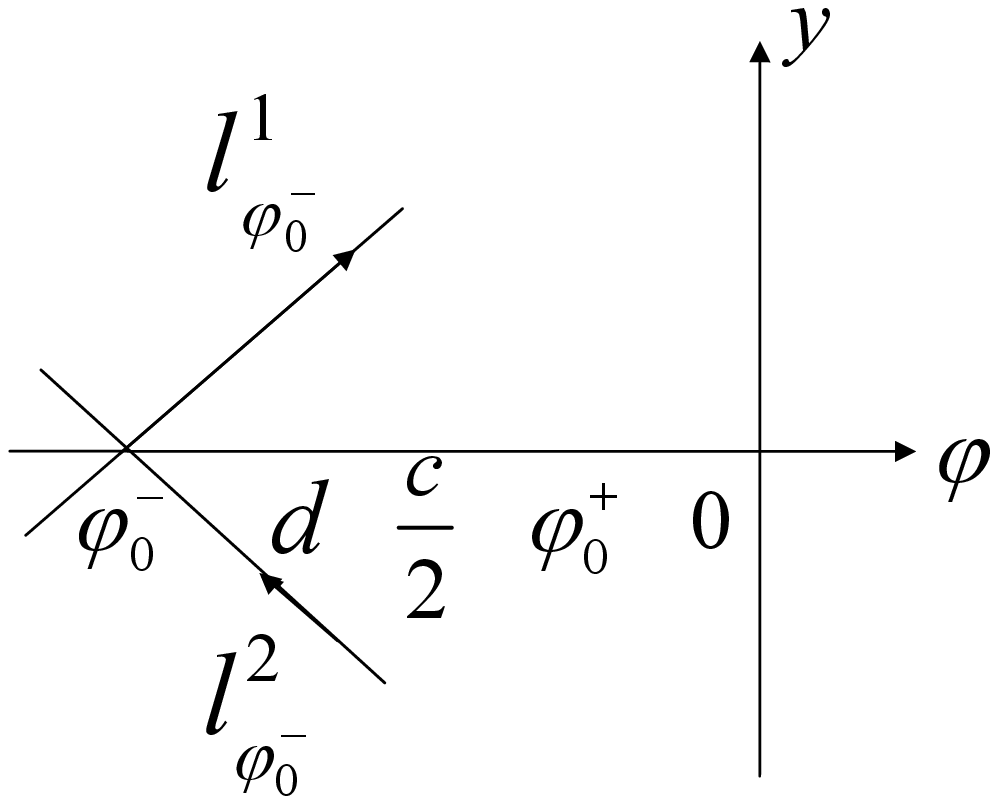}}\hspace{0.1\textwidth}
\subfloat[ ]{ \label{fig:4}
\includegraphics[height=1.6in,width=1.8in]{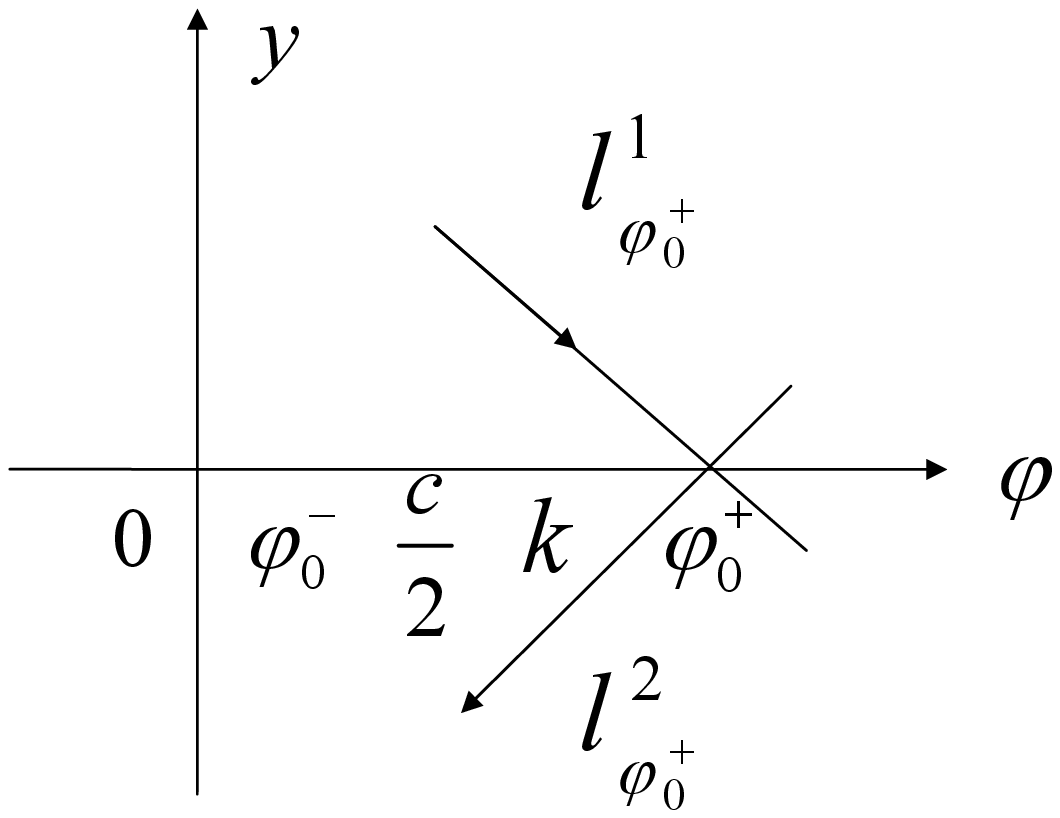}}\\

\caption{The sketches of orbits connecting with saddle points.
(a)$g>0$, $c<0$;
 (b)$g>0$, $c>0$; (c) $-\frac{2c^2}{9}<g\leq 0$, $c<0$; (d)$-\frac{2c^2}{9}<g\leq 0$, $c>0$.}
\end{figure}

On the $\varphi-y$ plane, the orbits $l_{\varphi _0^ - }^1 $,
$l_{\varphi _0^ - }^2 $, $l_{\varphi _0^ + }^1 $ and $l_{\varphi _0^
+ }^2 $ have the following expressions, respectively,
\begin{equation}
\label{eq4.1} l_{\varphi _0^ - }^1 :\quad y = \frac{(\varphi _0^ -
-\varphi)\sqrt {\varphi ^2 + l_1 \varphi + l_2 } }{2\varphi}
\end{equation}
\begin{equation}
\label{eq4.2} l_{\varphi _0^ - }^2 :\quad y = \frac{(\varphi
-\varphi _0^ - )\sqrt {\varphi ^2 + l_1 \varphi + l_2 } }{2\varphi}
\end{equation}
\begin{equation}
\label{eq4.3} l_{\varphi _0^ + }^1 :\quad y = \frac{(\varphi _0^ + -
\varphi )\sqrt {\varphi ^2 + m_1 \varphi + m_2 } }{2\varphi}
\end{equation}
\begin{equation}
\label{eq4.4} l_{\varphi _0^ + }^2 :\quad y = \frac{(\varphi -
\varphi _0^ + )\sqrt {\varphi ^2 + m_1 \varphi + m_2 } }{2\varphi}
\end{equation}
\noindent where $\varphi _0^ - $ and $\varphi _0^ + $ are in
(\ref{eq3.2}) , $l_1 $ and $l_2 $ are in (\ref{eq3.17}) and
(\ref{eq3.18}), $m_1 $ and $m_2 $ are in (\ref{eq3.20}) and
(\ref{eq3.21}), respectively.

Assume that $\varphi = \varphi _1 (\xi )$, $\varphi = \varphi _2
(\xi )$, $\varphi = \varphi _3 (\xi )$ and $\varphi = \varphi _4
(\xi )$ on $l_{\varphi _0^ - }^1 $, $l_{\varphi _0^ - }^2 $,
$l_{\varphi _0^ + }^1 $ and $l_{\varphi _0^ + }^2 $, respectively
and $\varphi _1 (0) = \varphi _2 (0) = a$, $\varphi _3 (0) = \varphi
_4 (0) = b$, $\varphi_2^* = \frac{c+\sqrt {c^2 +3g} }{2}$, where $a$
and $b$ are two constants satisfying $\varphi _0^ - <a <\frac{c}{2}$
and $\varphi_2^*<b<\varphi _0^ + $. Substituting
(\ref{eq4.1})-(\ref{eq4.4}) into the first equation of (\ref{eq2.4})
and integrating along the corresponding orbits, respectively, we
have
\begin{equation}
\label{eq4.5} \int_a^{\varphi _1 } {\frac{ - s}{(s - \varphi _0^ -
)\sqrt {s^2 + l_1 s + l_2 } }ds = \frac{1}{2}\int_0^\xi {ds} }
 \quad\quad(\textmd{along}  \quad l_{\varphi _0^ - }^1 ),
\end{equation}
\begin{equation}
\label{eq4.6} \int_{\varphi_2}^a {\frac{ s}{(s-\varphi _0^ -  )\sqrt
{s^2 + l_1 s + l_2 } }ds = \frac{1}{2}\int_\xi ^0 {ds} }
  \quad\quad(\textmd{along} \quad l_{\varphi _0^ - }^2 ),
\end{equation}
\begin{equation}
\label{eq4.7} \int_{\varphi _3 }^b {\frac{-s }{(s-\varphi _0^ +
)\sqrt {s^2 + m_1 s + m_2 } }ds = \frac{1}{2}\int_\xi ^0 {ds} }
 \quad(\textmd{along}  \quad l_{\varphi _0^ + }^1 ),
\end{equation}
\begin{equation}
\label{eq4.8} \int_b^{\varphi _4 } {\frac{s}{(s - \varphi _0^ +
)\sqrt {s^2 + m_1 s + m_2 } }ds = \frac{1}{2}\int_0^\xi {ds} }
 \quad(\textmd{along}  \quad l_{\varphi _0^ + }^2 ).
\end{equation}
With the aim of Maple, we obtain the implicit expressions of
$\varphi _i (\xi )$ as in (\ref{eq3.3})-(\ref{eq3.6}).

Meanwhile, suppose that $\varphi _1 (\xi ) \to \frac{c}{2}$ as $\xi
\to \xi _0^1 $, $\varphi _2 (\xi ) \to \frac{c}{2}$ as $\xi \to -
\xi _0^2 $, $\varphi _3 (\xi ) \to \varphi_2^*$ as $\xi \to - \xi
_0^3 $ and $\varphi _4 (\xi ) \to \varphi_2^*$ as $\xi \to \xi _0^4
$, then it follow from (\ref{eq4.5}) -(\ref{eq4.8}) that

\begin{equation}
\label{eq4.9} \xi _0^1 = \xi _0^2 = \int_a^{\frac{c}{2}} {\frac{ -
s}{(s - \varphi _0^ - )\sqrt {s^2 + l_1 s + l_2 } }ds}  \quad
\quad(\textmd{along}
 \quad l_{\varphi _0^ - }^1 ),
\end{equation}

\begin{equation}
\label{eq4.10} \xi _0^3 = \xi _0^4 = \int_b^{\varphi_2^*} {\frac{s
}{(s - \varphi _0^ + )\sqrt {s^2 + m_1 s + m_2 } }ds}  \quad
(\textmd{along} \quad l_{\varphi _0^ + }^2 ).
\end{equation}
With the aim of Maple, we get the expressions of $\xi _0^1 $ and
$\xi _0^3 $ as in (\ref{eq3.31}) and (\ref{eq3.32}). The proof of
(1) in Theorem (\ref{th3.1}) is finished.

Similarly, we can prove (2)-(4) in Theorem (\ref{th3.1}). Here we
omit the details.

\section{Numerical simulations}
\setcounter {equation}{0} In this section, we will simulate the
planar graphs of the kink-like and antikink-like wave solutions.

From Section 2, we see that in the parameter expressions $\varphi =
\varphi (\xi )$ and $y = y(\xi )$ of the orbits of system
(\ref{eq2.6}), the graph of $\varphi (\xi )$ and the integral curve
of Eq.(\ref{eq2.3}) are the same. In other words, the integral
curves of Eq.(\ref{eq2.3}) are the planar graphs of the traveling
waves of Eq.(\ref{eq1.7}). Therefore, we can see the planar graphs
of the kink-like and the antikink-like waves through the simulation
of the integral curves of Eq.(\ref{eq2.3}).

\begin{example}
 Take the same data as Example (\ref{ex3.1}), that is $g = 5$,
$c=-1$, $a =-0.75$, $b = 1.6$. Let $\varphi = a = -0.75$ in
(\ref{eq4.1}) and (\ref{eq4.2}), then we can get $y \approx 4.23397$
or $y \approx - 4.23397$. And let $\varphi = b = 1.6$ in
(\ref{eq4.3}) and (\ref{eq4.4}), then we obtain $y \approx 0.23198$
or $y \approx -0.23198$. Thus we take the initial conditions of
Eq.(\ref{eq2.3}) as follows: (a) Corresponding to $l_{\varphi _0^ -
}^1 $ we take $\varphi (0) = -0.75$ and $\varphi '(0) = 4.23397$;
(b) Corresponding to $l_{\varphi _0^ - }^2 $ we take $\varphi (0) =
-0.75$ and $\varphi '(0) = - 4.23397$; (c) Corresponding to
$l_{\varphi _0^ + }^1 $ we take $\varphi (0) = 1.6$ and $\varphi
'(0) = 0.23198$; (d) Corresponding to $l_{\varphi _0^ + }^2 $ we
take $\varphi (0) = 1.6$ and $\varphi '(0) = -0.23198$. Under each
set of initial conditions we use Maple to simulate the integrals
curve of Eq.(\ref{eq2.3}) in Fig.7.
\end{example}

\begin{figure}[h]
\centering \subfloat[]{\label{fig:1}
\includegraphics[height=1.8in,width=2.4in]{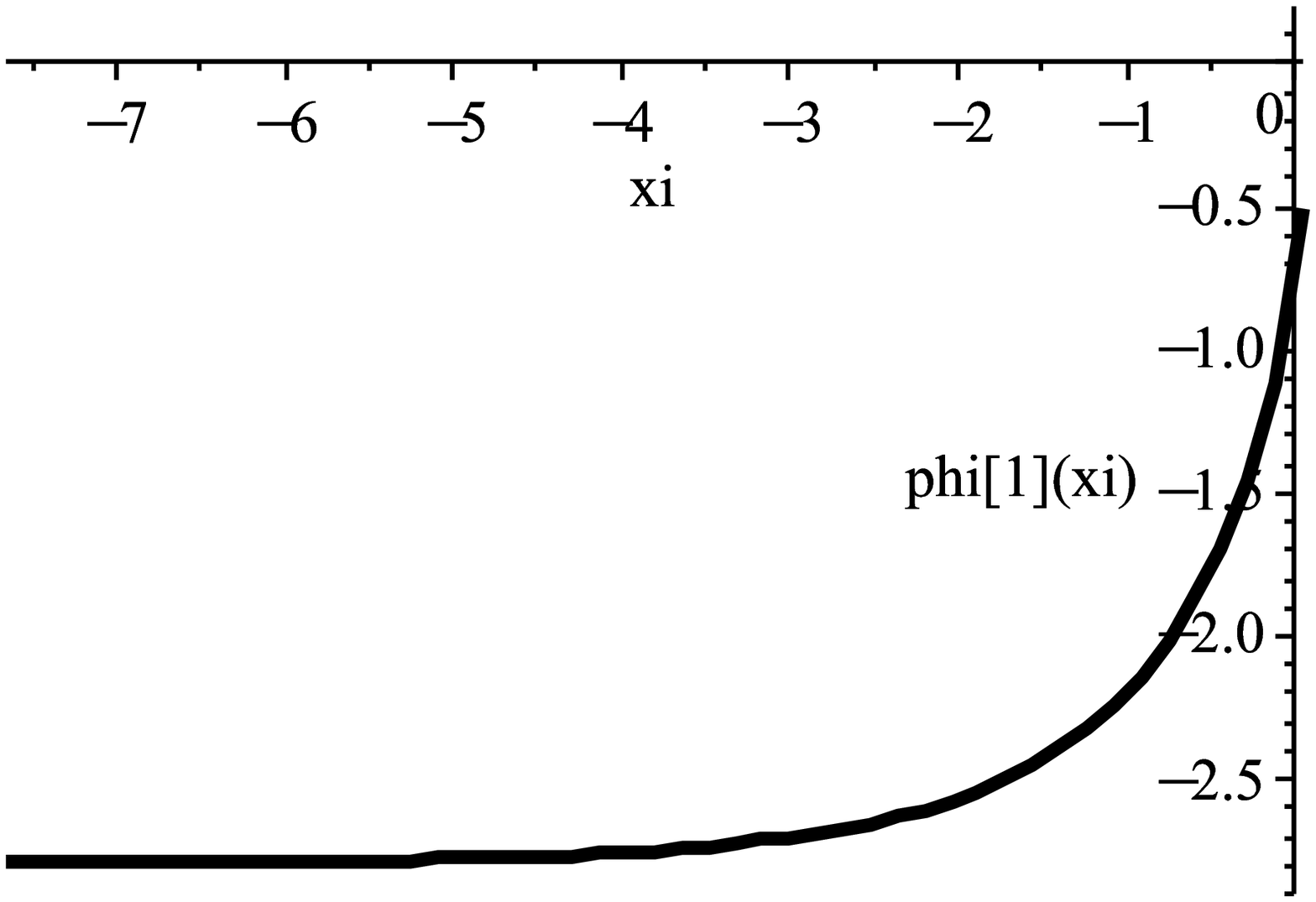}}\hspace{0.03\textwidth}
\subfloat[ ]{ \label{fig:2}
\includegraphics[height=1.8in,width=2.4in]{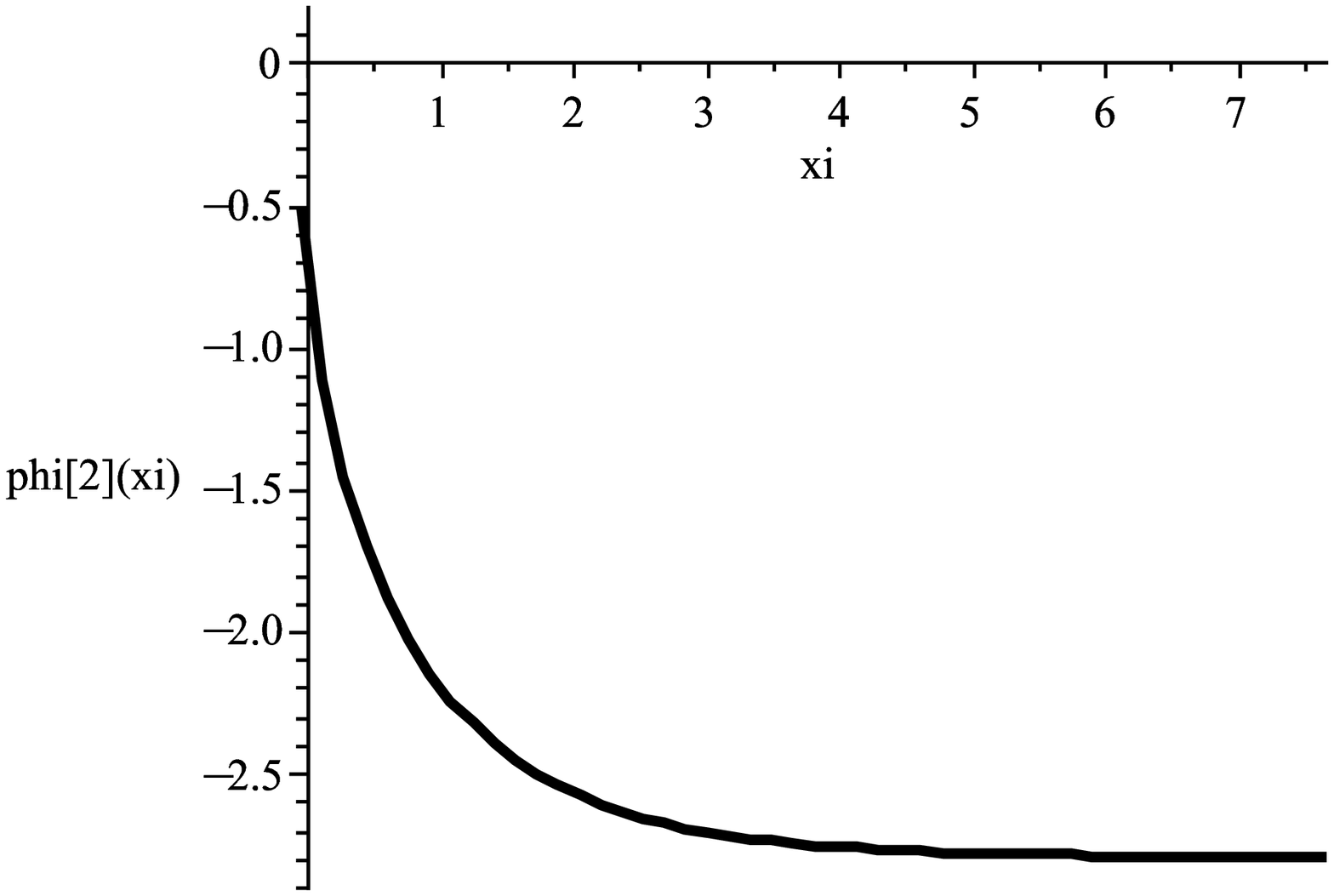}}\\
\subfloat[]{ \label{fig:3}
\includegraphics[height=1.8in,width=2.4in]{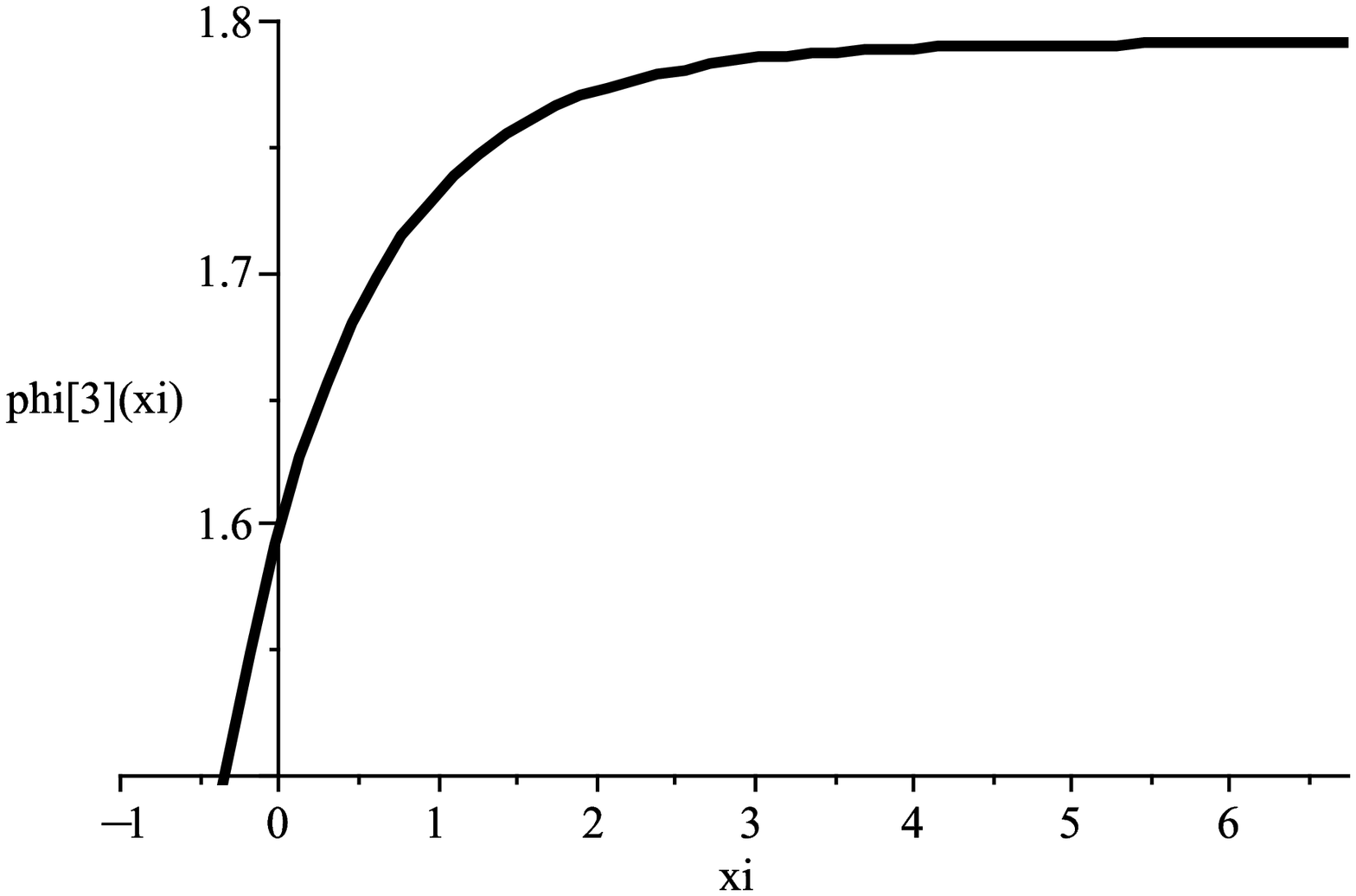}}\hspace{0.05\textwidth}
\subfloat[ ]{ \label{fig:4}
\includegraphics[height=1.8in,width=2.4in]{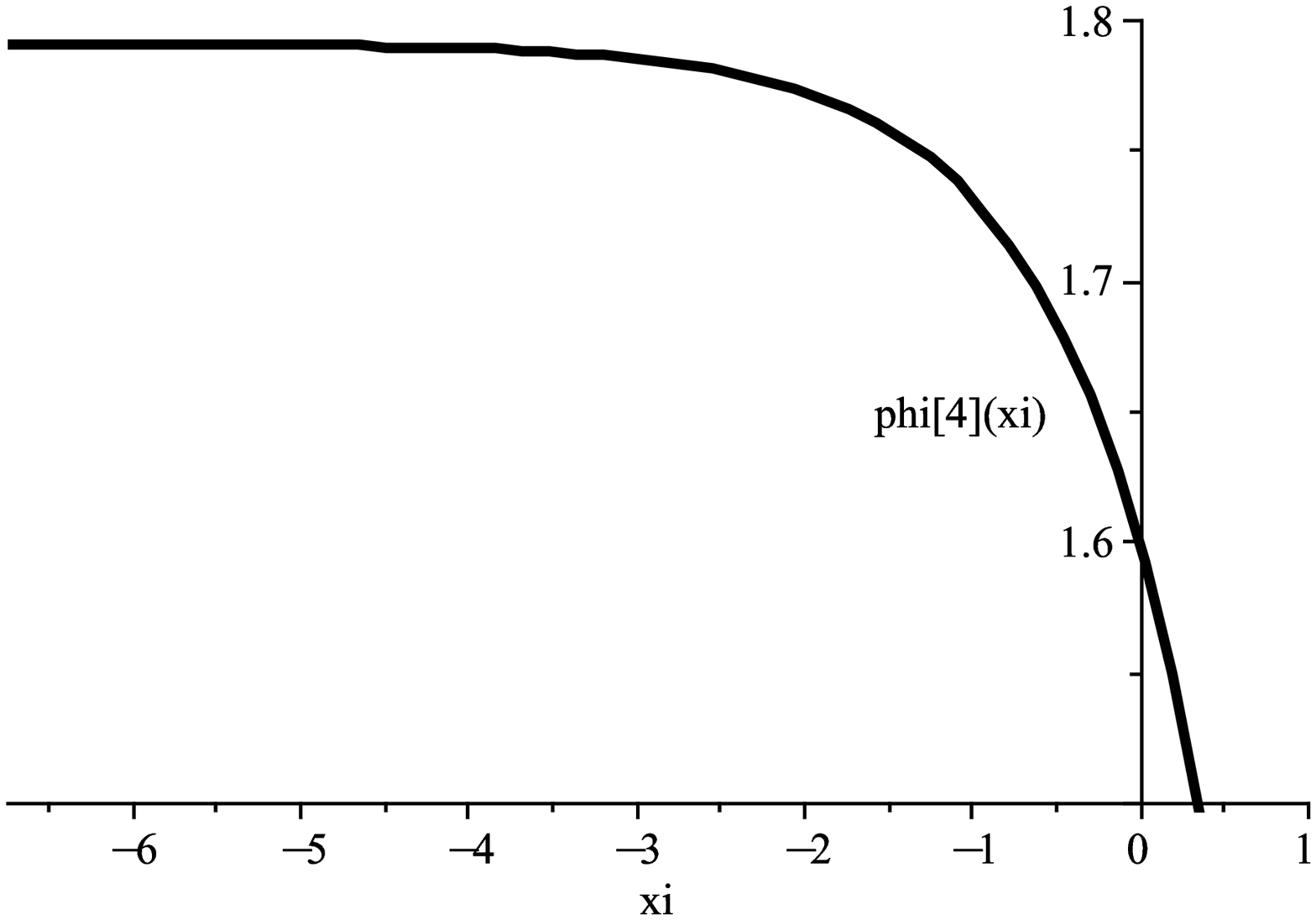}}\\
\caption{The numerical simulations of integral curves of
Eq.(\ref{eq2.3}) when $g=5$ and $c=-1$. (a) $\varphi(0)=-0.75$,
$\varphi'(0)=4.23397$; (b) $\varphi(0)=-0.75$,
$\varphi'(0)=-4.23397$; (c) $\varphi(0)=1.6$,
$\varphi'(0)=0.230189$; (d) $\varphi(0)=1.6$,
$\varphi'(0)=-0.230189$.}
\end{figure}

\begin{example}
 Take the same data as Example (\ref{ex3.2}), that is $g =5$,
$c = 1$, $a = -1.6$, $b = 2$. Let $\varphi = a =-1.6$ in
(\ref{eq4.1}) and (\ref{eq4.2}) , then we can get $y \approx
0.230189$ or $y \approx - 0.230189$. And let $\varphi = b = 2$ in
(\ref{eq4.3}) and (\ref{eq4.4}), then we obtain $y \approx 0.849228$
or $y \approx-0.849228$. Thus we take the initial conditions of
Eq.(\ref{eq2.3}) as follows: (a) Corresponding to $l_{\varphi _0^ -
}^1 $ we take $\varphi (0) = -1.6$ and $\varphi '(0) = 0.230189$;
(b) Corresponding to $l_{\varphi _0^ - }^2 $ we take $\varphi (0) =
-1.6$ and $\varphi '(0) = -0.230189$; (c) Corresponding to
$l_{\varphi _0^ + }^1 $ we take $\varphi (0) = 2$ and $\varphi '(0)
= 0.849228$; (d) Corresponding to $l_{\varphi _0^ + }^2 $ we take
$\varphi (0) = 2$ and $\varphi '(0) = -0.849228$. Under each set of
initial conditions we use Maple to simulate the integrals curve of
Eq.(\ref{eq2.3}) in Fig.8.
\end{example}

\begin{figure}[h]
\centering \subfloat[]{\label{fig:1}
\includegraphics[height=1.7in,width=2.3in]{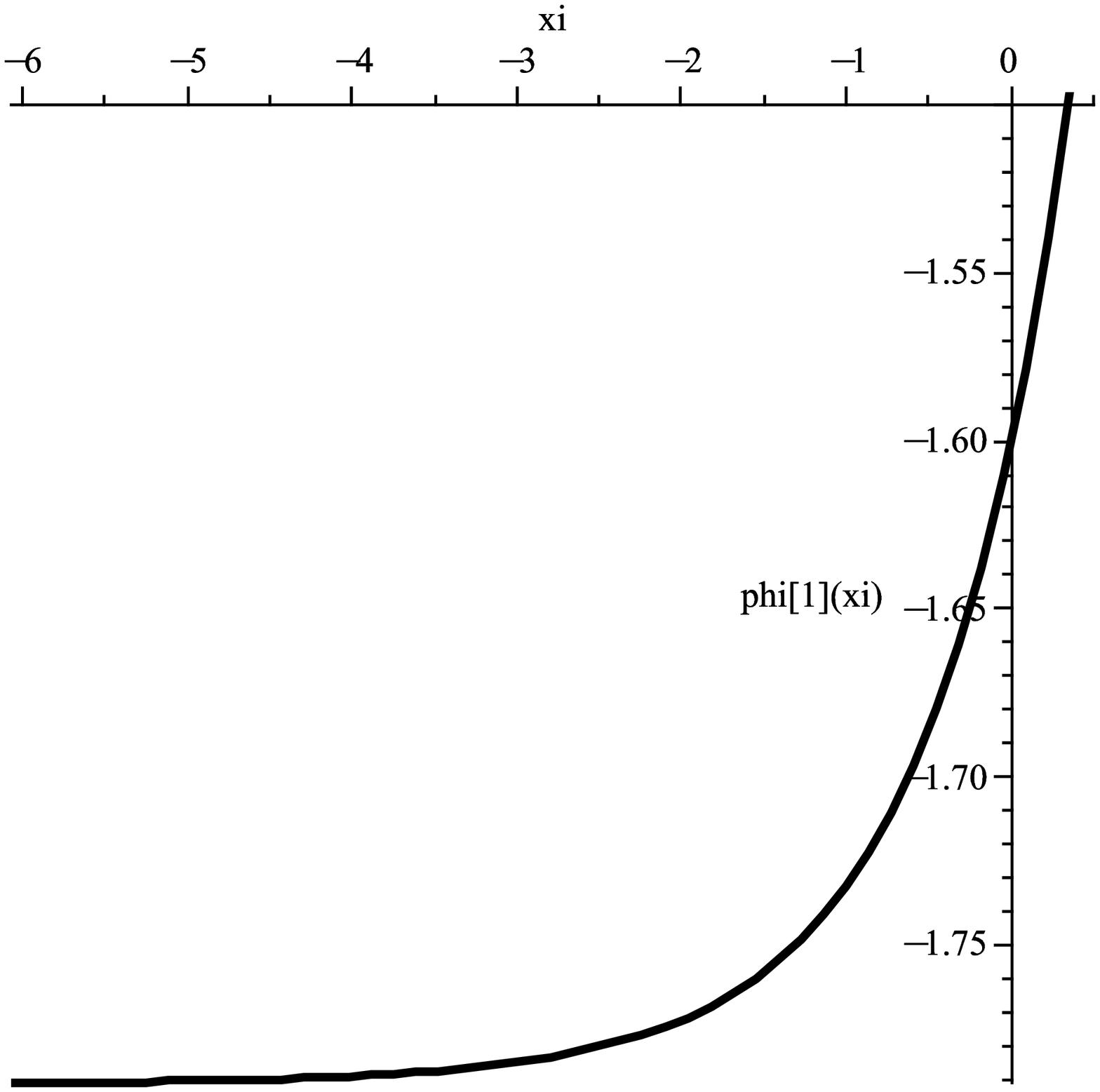}}\hspace{0.04\textwidth}
\subfloat[ ]{ \label{fig:2}
\includegraphics[height=1.8in,width=2.4in]{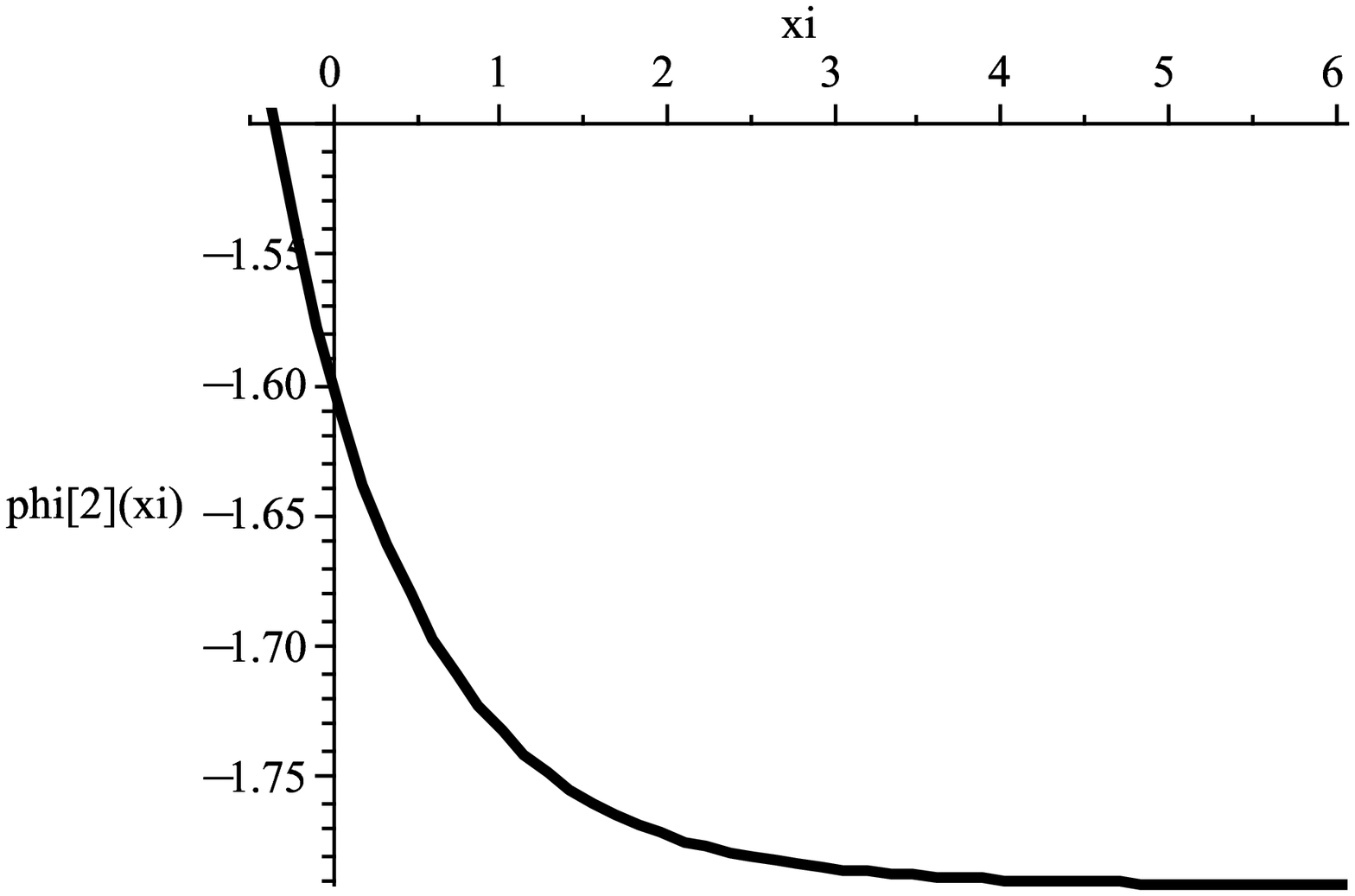}}\\
\subfloat[]{ \label{fig:3}
\includegraphics[height=1.8in,width=2.4in]{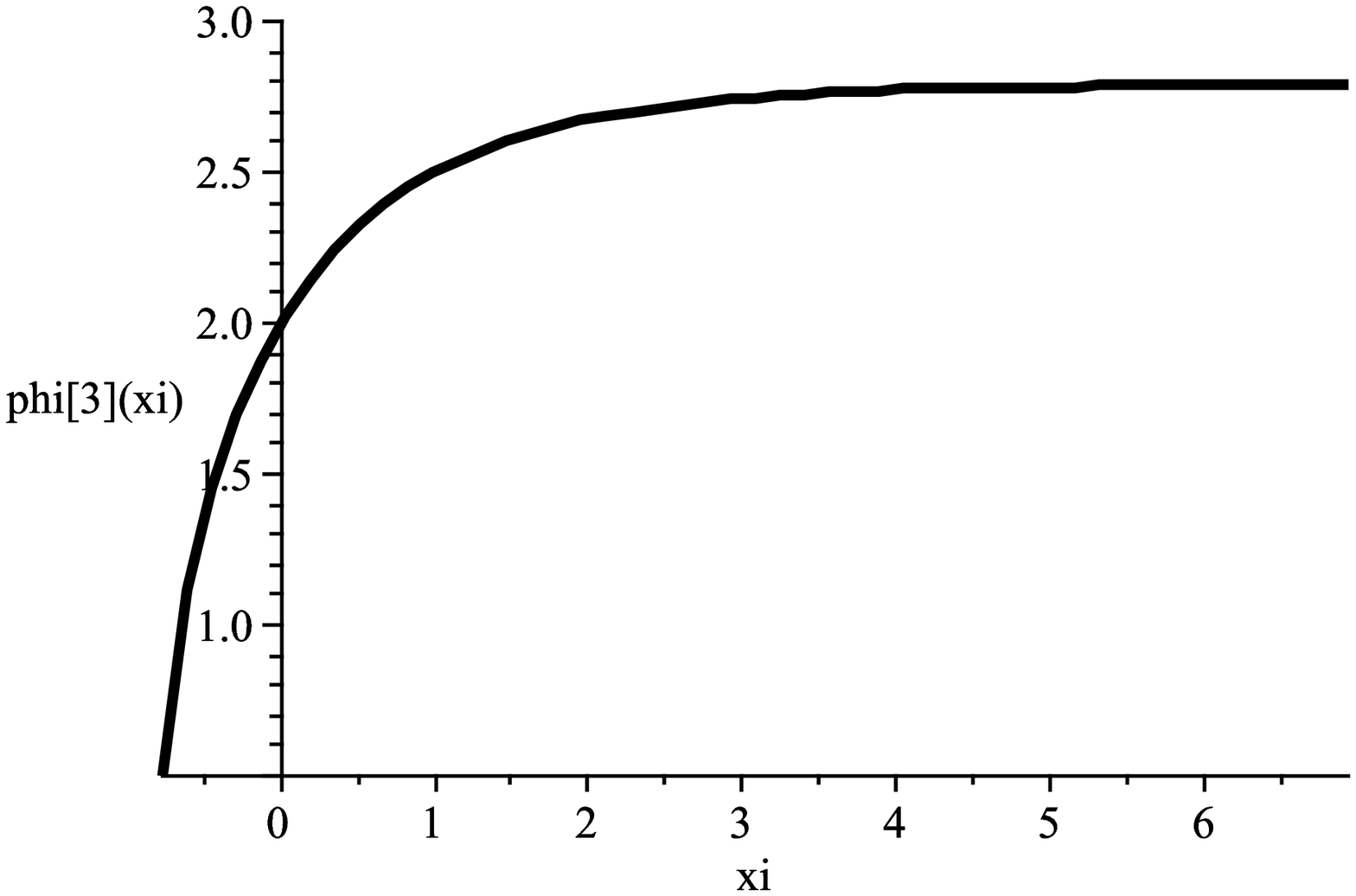}}\hspace{0.06\textwidth}
\subfloat[ ]{ \label{fig:4}
\includegraphics[height=1.8in,width=2.3in]{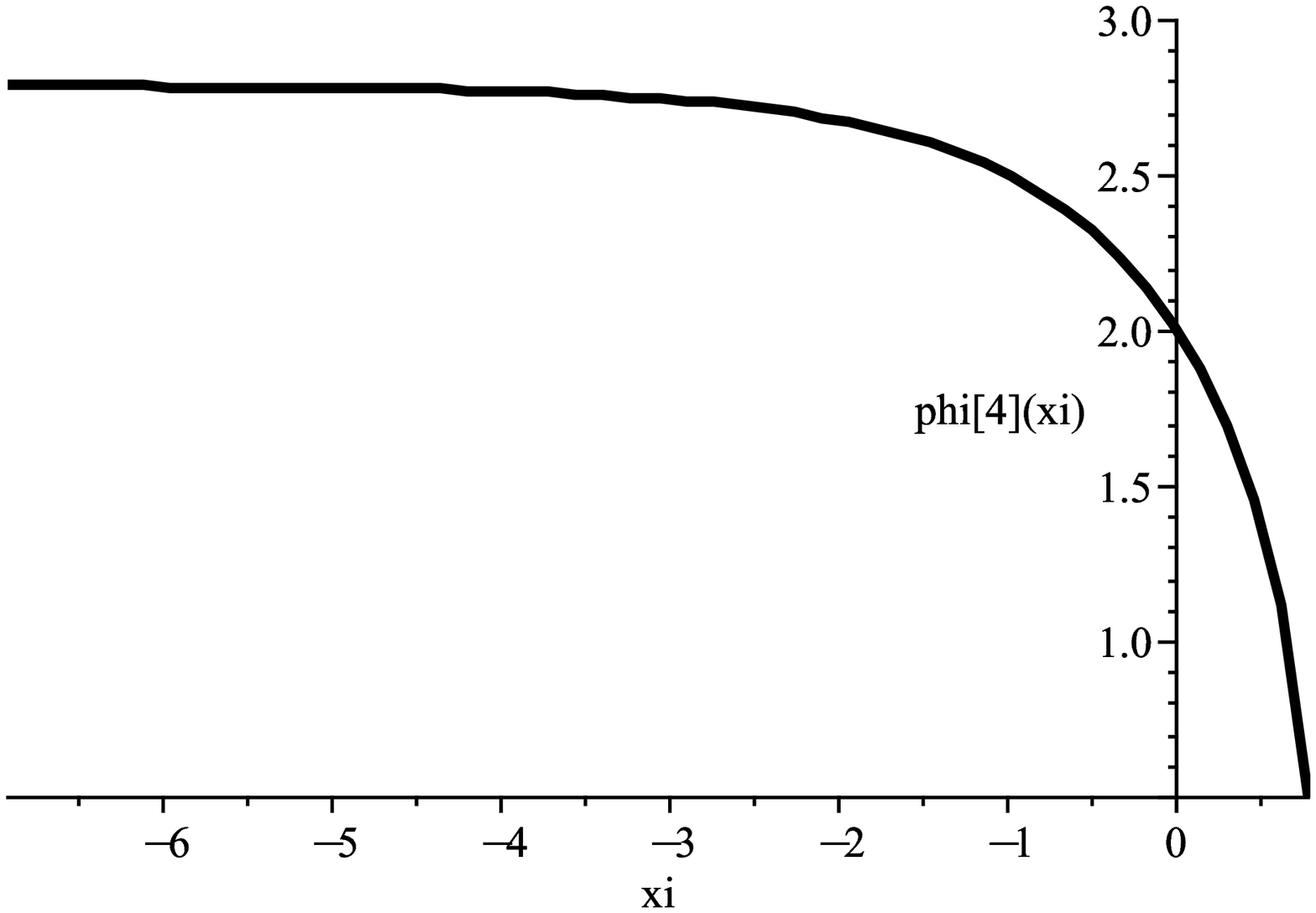}}\\
\caption{The numerical simulations of integral curves of
Eq.(\ref{eq2.3}) when $g=5$ and $c=1$. (a) $\varphi(0)=-1.6$,
$\varphi'(0)=0.230189$; (b) $\varphi(0)=-1.6$,
$\varphi'(0)=-0.230189$; (c) $\varphi(0)=2$, $\varphi'(0)=0.849228$;
(d) $\varphi(0)=2$, $\varphi'(0)=-0.849228$.}
\end{figure}

\begin{example} Take the same data as Example (\ref{ex3.3}), that is $g = -0.5$, $c = -2$,
 $d =-1.2$. Let $\varphi = d = -1.2$ in (\ref{eq4.1}) and
(\ref{eq4.2}), then we can get $y \approx 0.364453$ or $y \approx -
0.364453$. Thus we take the initial conditions of Eq.(\ref{eq2.3})
as follows: (a) Corresponding to $l_{\varphi _0^ - }^{1} $ we take
$\varphi (0) = -1.2$ and $\varphi '(0) = 0.364453$; (b)
Corresponding to $l_{\varphi _0^ - }^{2} $ we take $\varphi (0) =
-1.2$ and $\varphi '(0) = - 0.364453$. Under each set of initial
conditions we use Maple to simulate the integrals curve of
Eq.(\ref{eq2.3}) in Fig.9.
\end{example}

\begin{figure}[h]
\centering \subfloat[]{\label{fig:1}
\includegraphics[height=1.8in,width=2.4in]{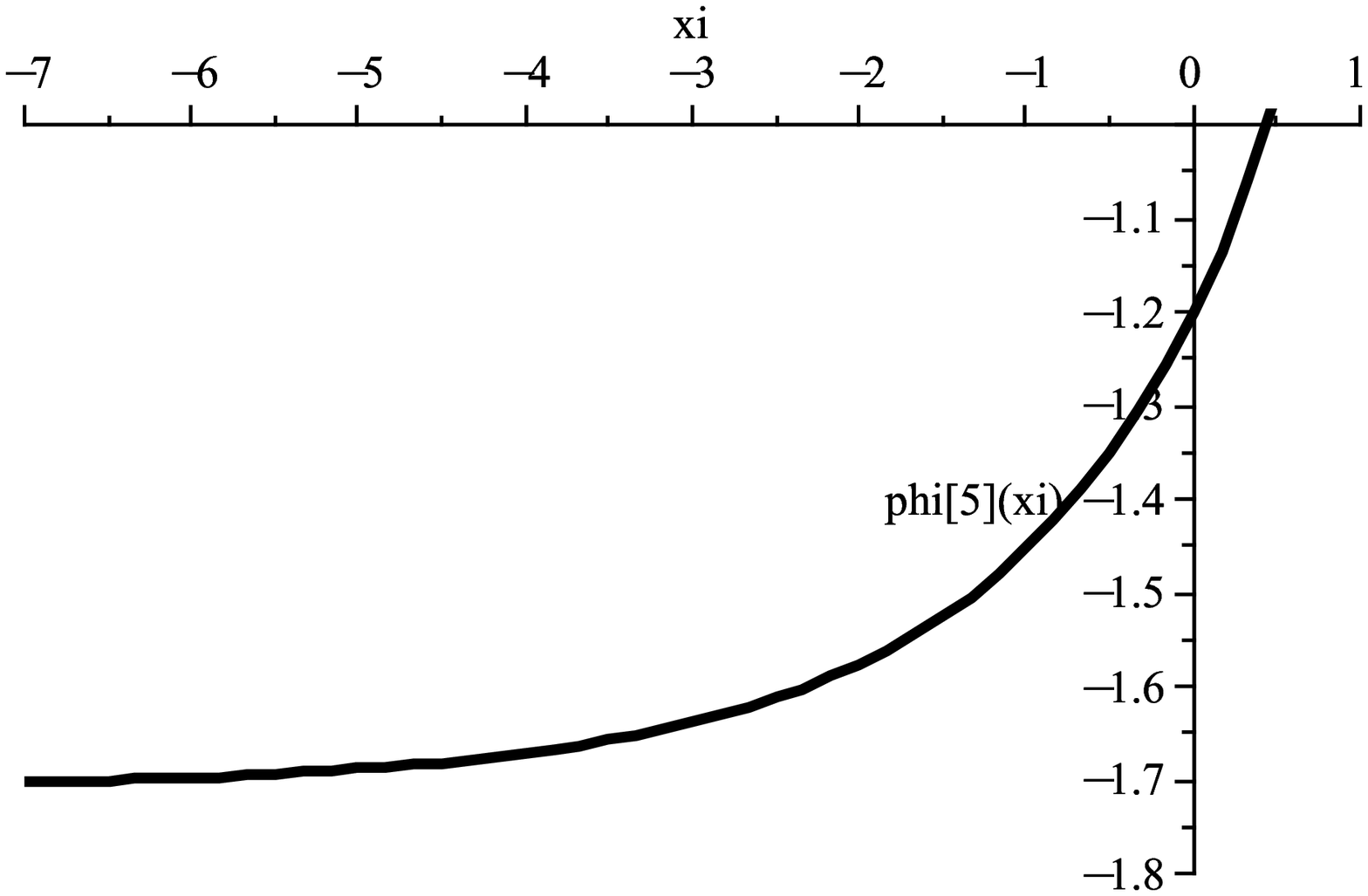}}\hspace{0.03\textwidth}
\subfloat[ ]{ \label{fig:2}
\includegraphics[height=1.7in,width=2.4in]{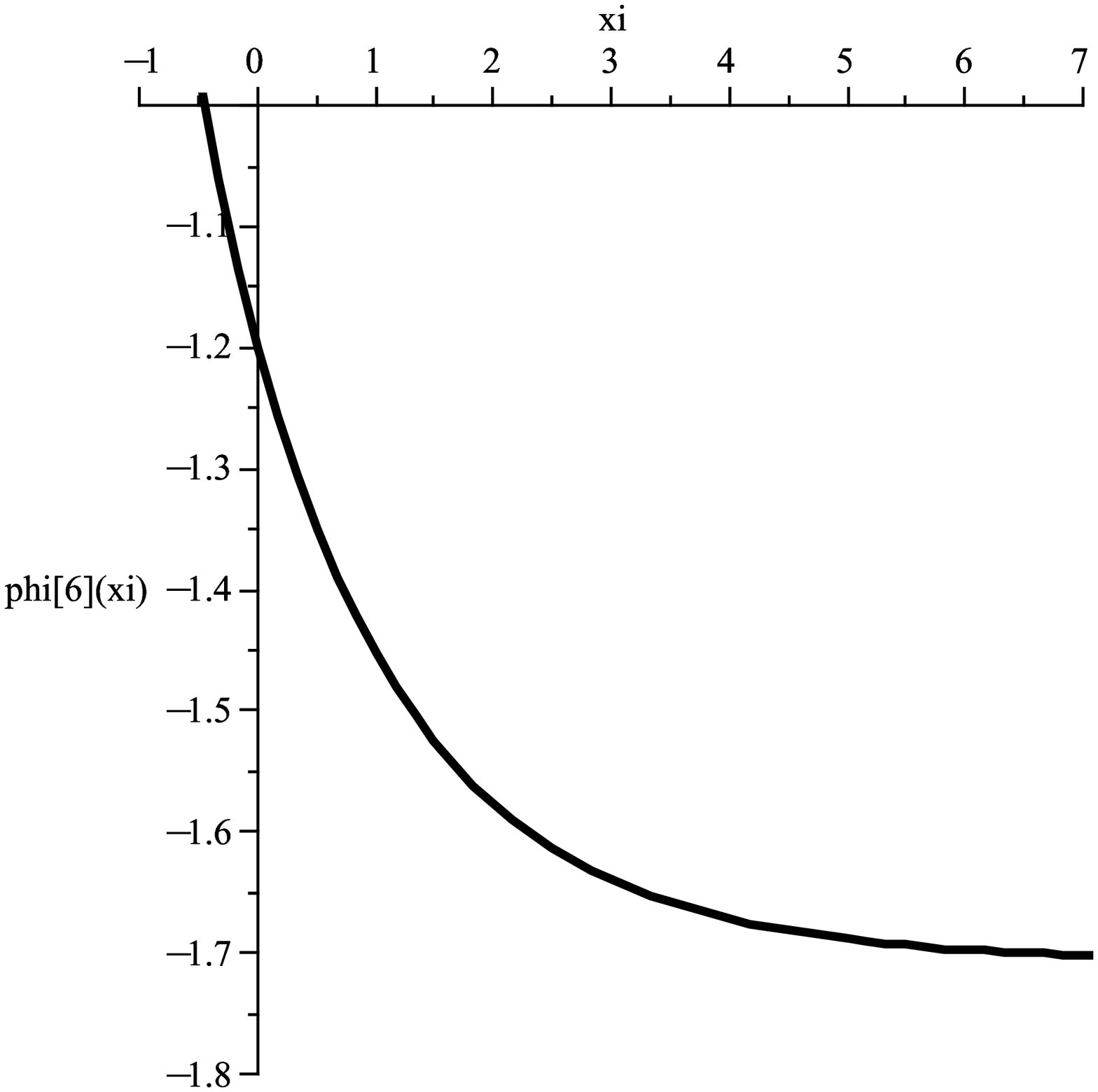}}\\
\caption{The numerical simulations of integral curves of
Eq.(\ref{eq2.3}) when $g=-0.5$ and $c=-2$. (a) $\varphi(0)=-1.6$,
$\varphi'(0)=0.230189$; (b) $\varphi(0)=-1.6$,
$\varphi'(0)=-0.230189$.}
\end{figure}

\begin{example}
Take the same data as Example (\ref{ex3.4}), that is $g =-0.5$, $c =
2$, $k = 1.2$. Let $\varphi = k =-1.6$ in (\ref{eq4.3}) and
(\ref{eq4.4}) , then we can get $y \approx 0.364453$ or $y \approx -
0.364453$. Thus we take the initial conditions of Eq.(\ref{eq2.3})
as follows: (a) Corresponding to $l_{\varphi _0^ + }^{1} $ we take
$\varphi (0) = 1.2$ and $\varphi '(0) = 0.364453$; (b) Corresponding
to $l_{\varphi _0^ + }^{2} $ we take $\varphi (0) = 1.2$ and
$\varphi '(0) = -0.364453$. Under each set of initial conditions we
use Maple to simulate the integrals curve of Eq.(\ref{eq2.3}) in
Fig.10.
\end{example}

\begin{figure}[h]
\centering \subfloat[]{\label{fig:1}
\includegraphics[height=1.8in,width=2.4in]{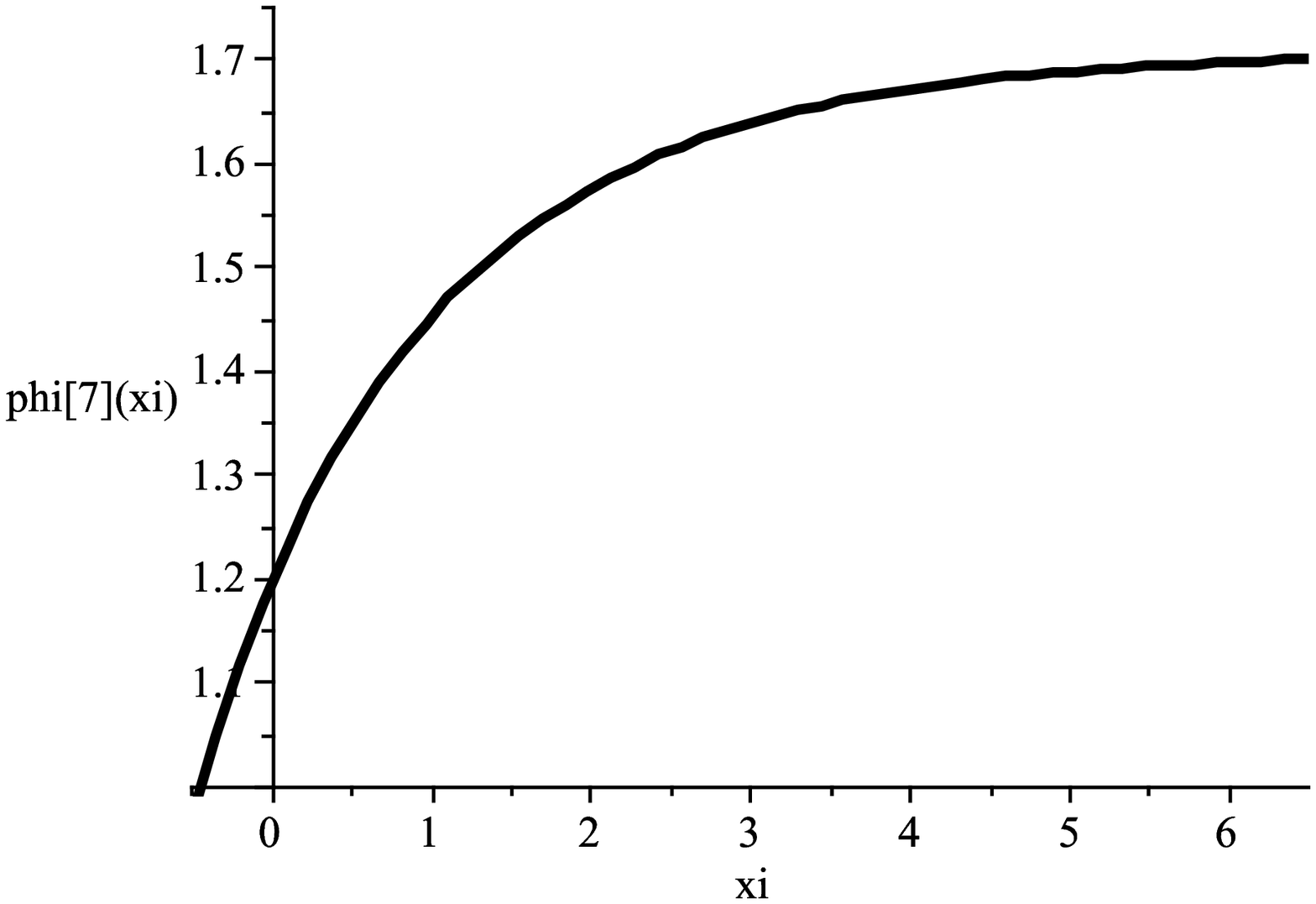}}\hspace{0.04\textwidth}
\subfloat[ ]{ \label{fig:2}
\includegraphics[height=1.8in,width=2.4in]{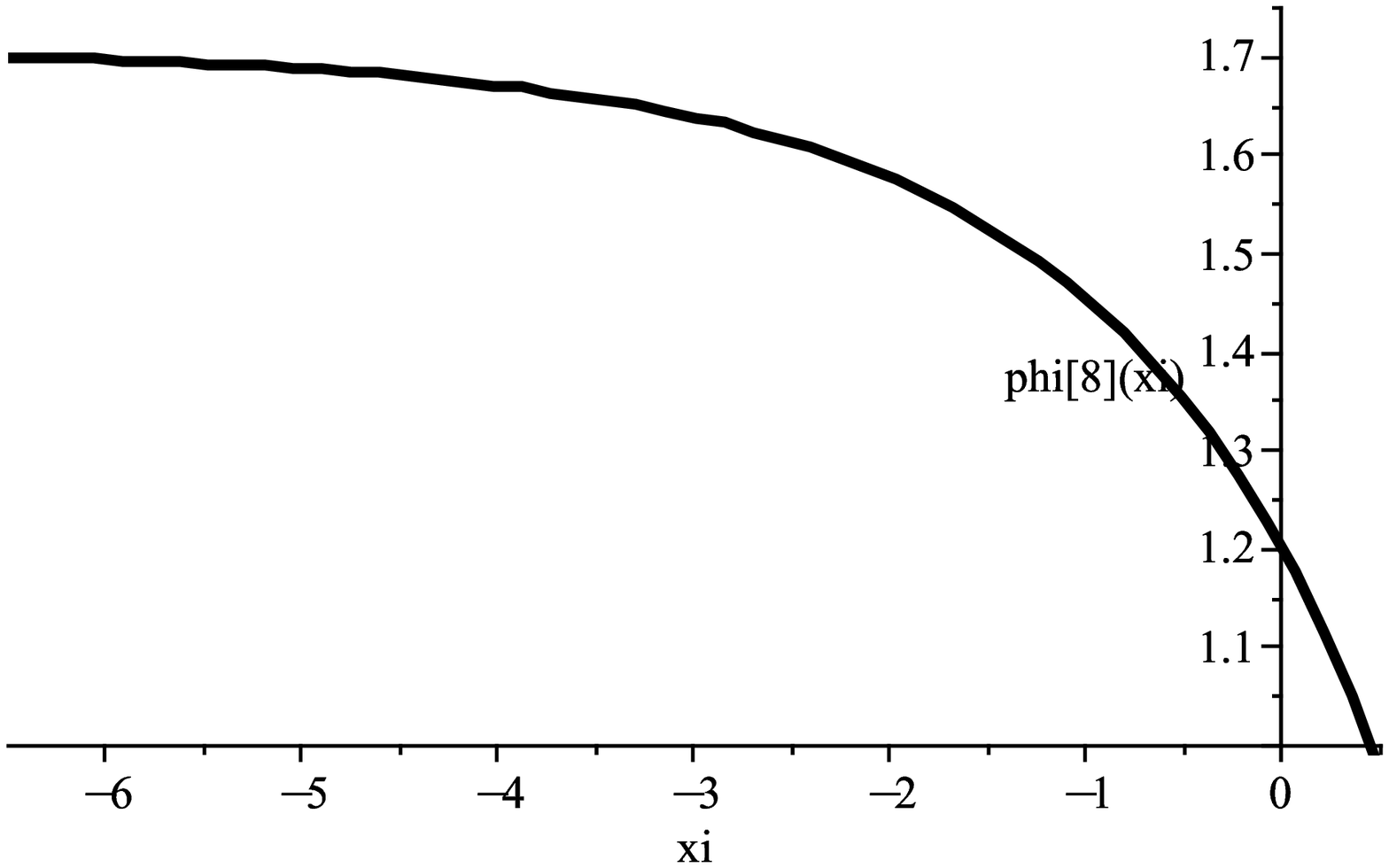}}\\
\caption{The numerical simulations of integral curves of
Eq.(\ref{eq2.3}) when $g=-0.5$ and $c=2$. (a) $\varphi(0)=1.2$,
$\varphi'(0)=0.364453$; (b) $\varphi(0)=1.2$,
$\varphi'(0)=-0.364453$.}
\end{figure}

Comparing Fig.2 with Fig.7,  Fig.3 with Fig.8, Fig.4 with Fig.9, and
Fig.5 with Fig.10, we can see that the graphs of
$\varphi_i(\xi)(i=1,2,3,4,5,6,7,8)$ are the same as the
corresponding integral curves of Eq.(\ref{eq2.3}). This implies that
our theoretical results agree with the numerical simulations.

\section{Conclusion}

In this paper, we find a new type of bounded travelling wave
solutions for the $K(2,2)$ equation with osmosis dispersion. Their
implicit expressions are obtained in (\ref{eq3.3})-(\ref{eq3.14}).
From the graphs (see Fig.2-Fig.5) of the implicit functions and the
numerical simulations (see Fig.7-Fig.10) we see that these new
bounded solutions are defined on some semifinal bounded domains and
possess properties of kink and anti-kink wave solutions.

\end{document}